\documentclass[reprint,prl,aps]{revtex4-1}
\usepackage{graphicx}
\usepackage{color}
\usepackage{dcolumn}
\usepackage{bm}
\usepackage{amssymb}
\usepackage{color,amsmath}
\usepackage{soul,xcolor} 
\setstcolor{red} 
\usepackage{epstopdf}
\usepackage{lipsum}

\begin{document}

\title{\large {Unraveling intrinsic flexoelectricity in twisted double bilayer graphene}}

\author{Yuhao Li$^{1}$}
\author{Xiao Wang$^{2,3}$}
\author{Deqi Tang$^{3}$}
\author{Xi Wang$^{1,4}$}
\author{K. Watanabe$^{5}$}
\author{T. Taniguchi$^{6}$} 
\author{Daniel R. Gamelin$^{4}$} 
\author{David H. Cobden$^{1}$}
\author{Matthew Yankowitz$^{1,7}$}
\author{Xiaodong Xu$^{1,7,*}$}
\author{Jiangyu Li$^{2,3,8,*}$}

\affiliation{$^{1}$Department of Physics, University of Washington, Seattle, WA 98195, USA}
\affiliation{$^{2}$Department of Materials Science and Engineering, Southern University of Science and Technology, Shenzhen 518055, Guangdong, China}
\affiliation{$^{3}$Shenzhen Key Laboratory of Nanobiomechanics, Shenzhen Institutes of Advanced Technology, Chinese Academy of Sciences, Shenzhen 518055, Guangdong, China}
\affiliation{$^{4}$Department of Chemistry, University of Washington, Seattle, WA 98195, USA}
\affiliation{$^{5}$Research Center for Functional Materials, National Institute for Materials Science, 1-1 Namiki, Tsukuba 305-0044, Japan}
\affiliation{$^{6}$International Center for Materials Nanoarchitectonics, National Institute for Materials Science, 1-1 Namiki, Tsukuba 305-0044, Japann}
\affiliation{$^{7}$Department of Materials Science and Engineering, University of Washington, Seattle, WA 98195, USA}
\affiliation{$^{8}$Guangdong Key Provisional Laboratory of Functional Oxide Materials and Devices, Southern University of Science and Technology, Shenzhen 518055, Guangdong, China}
\affiliation{$^{*}$lijy@sustech.edu.cn(J.L.); xuxd@uw.edu(X.X.)}


\begin{abstract}
\noindent Moiré superlattices of two-dimensional (2D) materials with a small twist angle are thought to exhibit appreciable flexoelectric effect, though unambiguous confirmation of their flexoelectricity is challenging due to artifacts associated with commonly used piezoresponse force microscopy (PFM). For example, unexpectedly small phase contrast ($\sim$$8^{\circ}$) between opposite flexoelectric polarizations was reported in twisted bilayer graphene (tBG), though theoretically predicted value is $180^{\circ}$. Here we developed a methodology to extract intrinsic moiré flexoelectricity using twisted double bilayer graphene (tDBG) as a model system, probed by lateral PFM. For small twist angle samples, we found that a vectorial decomposition is essential to recover the small intrinsic flexoelectric response at domain walls from a large background signal. The obtained three-fold symmetry of commensurate domains with significant flexoelectric response at domain walls is fully consistent with our theoretical calculations. Incommensurate domains in tDBG with relatively large twist angles can also be observed by this technique. Our work provides a general strategy for unraveling intrinsic flexoelectricity in van der Waals moiré superlattices while providing insights into engineered symmetry breaking in centrosymmetric materials.
\\ \hspace*{\fill} 
\end{abstract}

\maketitle

\begin{figure}[htbp]	  
	  \centering
	  \includegraphics[width=3.2in] {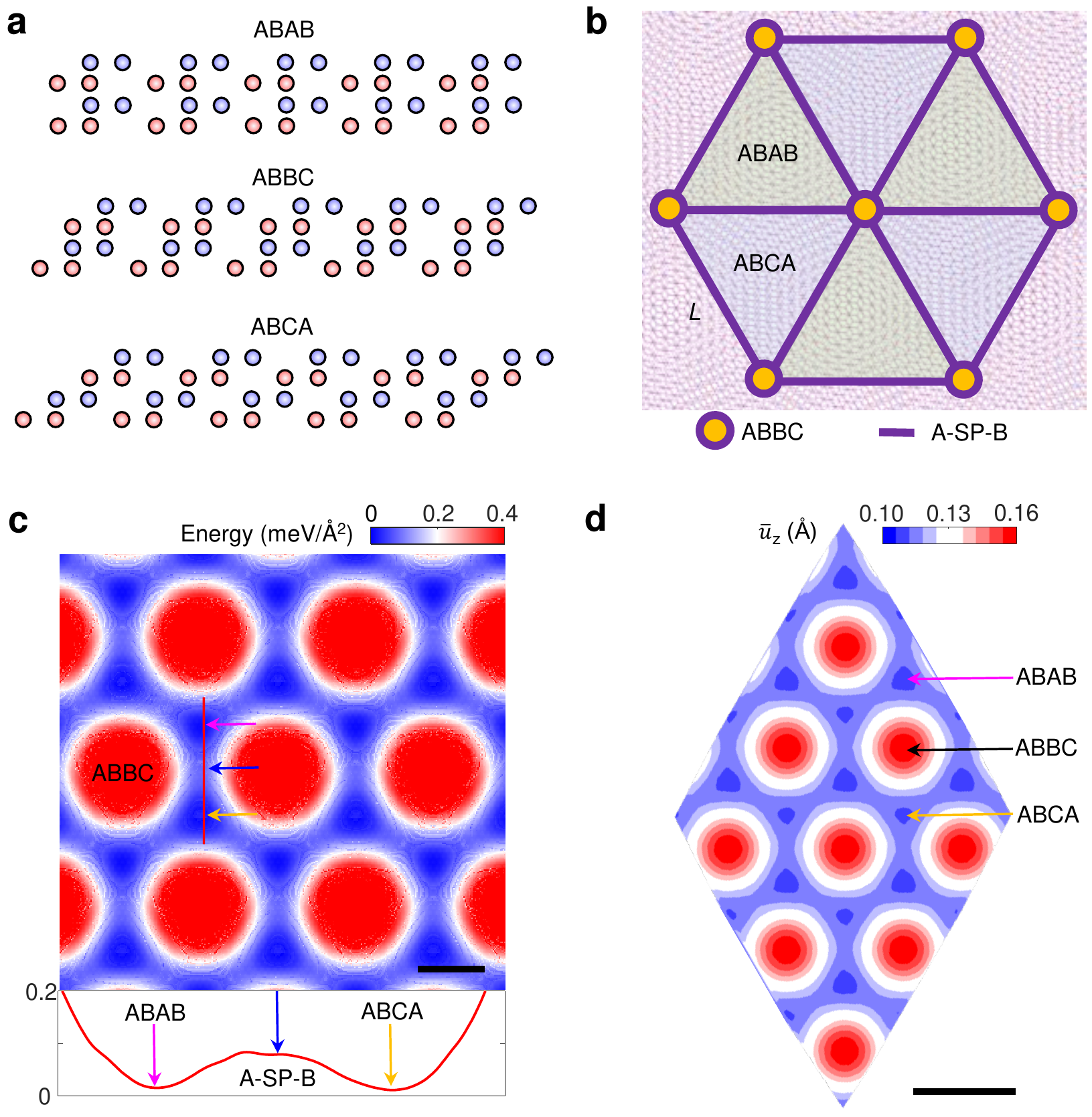}
	  \caption
	  { \textbf{Atomic stacking and density functional theory calculation of twisted double bilayer graphene.}
	 	\textbf{(a)} Side view of three atomic stacking domains (ABAB, ABBC and ABAB) in a tDBG moiré superlattice.
	 	\textbf{(b)} Schematic of tDBG moiré superlattice and relative locations of three domains. By rotating two bilayer graphene with a small angle $\theta$, a moiré superlattice is formed that is characterized by its wavelength $L=(a/2)\cdot\text{csc}(\theta/2)$, which is related to the lattice constant of graphene, $a$.
	 	\textbf{(c)} PES between two Bernal-stacked (AB) graphene bilayers. The ABAB, A-SP-B and ABCA regions are marked with pink, blue and orange arrows, respectively. The energy is normalized by the area of the unit cell. The profile of energy marked using the red line is presented and the corresponding domains are given. Scale bar, 1 Å.
	 	\textbf{(d)} Average out-of-plane displacement field $\overline{u}_{z}$ of the reconstructed tDBG moiré superlattice. The ABAB, ABBC and ABCA regions are marked with pink, black and orange arrows, respectively. Scale bar, 2 nm.
 } 
	 \label{fig:1} 
\end{figure}

\section*{Introduction}
Moiré superlattices in two-dimensional (2D) van der Waals (vdW) heterostructures, which consist of layers with a small twist angle, have provided a flexible and versatile platform for observing correlated electron phenomena\cite{RN1603}. Recently, twisted bilayer graphene (tBG) at the so-called magic angle ($\sim$$1.1^{\circ}$) has been reported to exhibit a variety of novel physical properties from superconductivity\cite{RN989}, ferromagnetism\cite{RN1515, RN1780}, ferroelectricity\cite{RN1785} to flexoelectricity\cite{RN1679}. Similar correlated states with a number of emergent properties also hold in twisted double bilayer graphene (tDBG)\cite{RN1721, RN1695, RN1562}. Of particular interest here is flexoelectric polarization arising from a strain gradient, which is a universal property exhibited by materials of all symmetries\cite{RN2142, RN2143, RN2141, RN1575}, though it only becomes prominent at the nanoscale. The unambiguous confirmation of the flexoelectric effect, however, is challenging, since piezoresponse force microscopy (PFM) most conveniently used for its characterization is prone to artifacts\cite{RN1773, RN137} and often yield inconclusive data. Indeed, only small PFM phase contrast is observed in tBG\cite{RN1679}, while $180^{\circ}$ is expected, and a complete theoretical understanding of the flexoelectric effect in moiré superlattices is still lacking. In this regard, the demonstration of moiré superlattice induced flexoelectricity would be a powerful addition to traditional piezoelectricity, serving as a beacon for symmetry-breaking engineering of centrosymmetric materials\cite{RN2164, RN1575, RN2145, RN2152}.

Here, we calculate potential energy surfaces (PESs) for two sheets of Bernal-stacked (AB) bilayer graphene. We optimize the corresponding tDBG moiré superlattice with twist angle of $6.01^{\circ}$ using density functional theory (DFT), revealing its three-fold symmetry reduced from unoptimized six-fold one due to the incompatibility of different domains. These atomistic calculations are then complemented by large scale finite element analysis (FEA) to account for the effect of lattice mismatch, demonstrating three-fold symmetric in-plane flexoelectric polarization across domain walls of tDBG\cite{RN1662}. To confirm this prediction, we fabricate tDBG samples with twist angles of $0.1^{\circ}$$\sim$$0.5^{\circ}$ and measure their in-plane electromechanical responses using lateral piezoelectric force microscopy (LPFM). Next, we develop a vectorial analysis to decouple the intrinsic flexoelectric response from background noise\cite{RN1689}, recovering the expected three-fold symmetry of commensurate domains, fully consistent with theoretical expectation. Incommensurate domains in tDBG with larger twist angles can also be observed by this technique.

\begin{figure*}[htbp]	  
	\centering
	\includegraphics[width=6in] {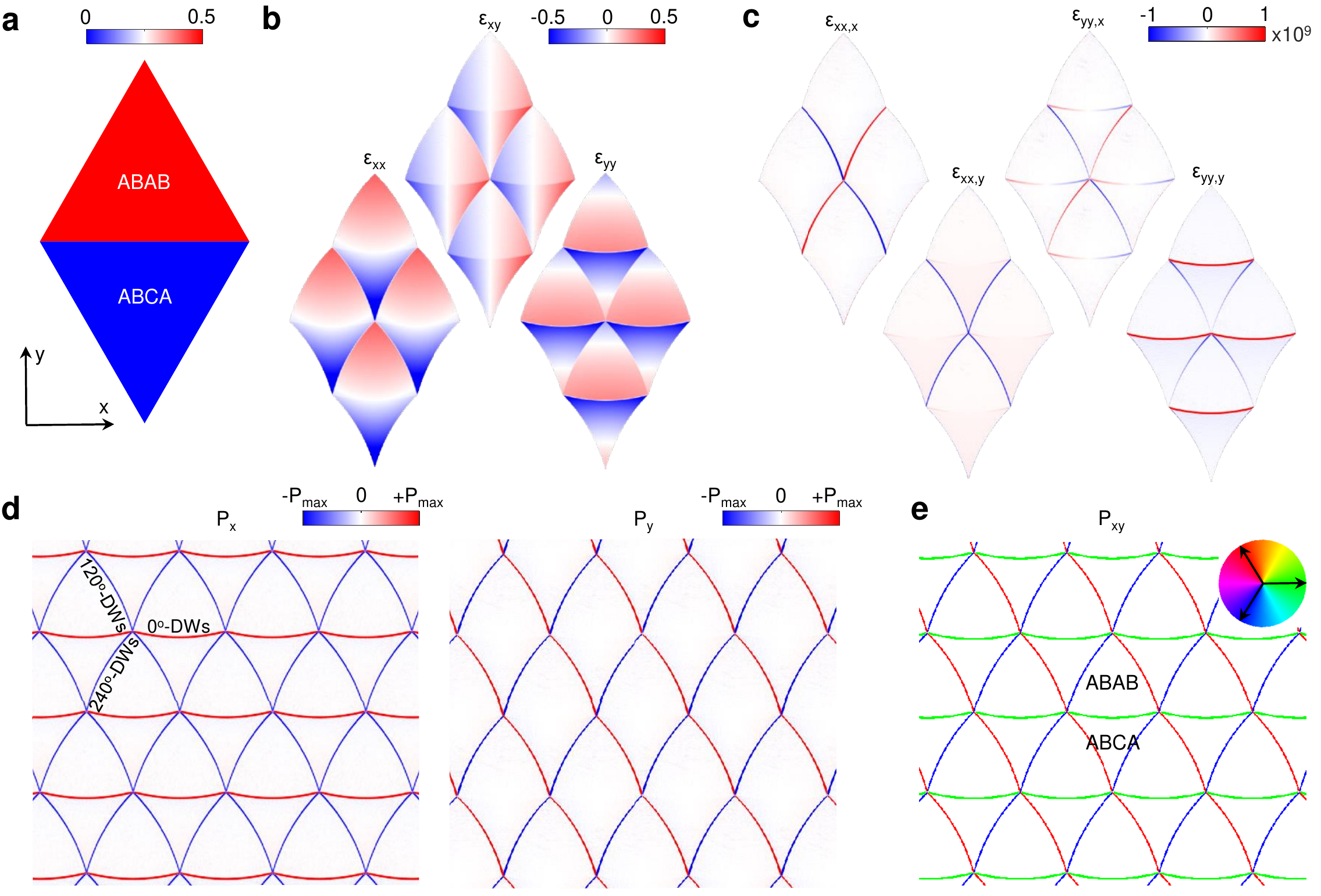}
	\caption
	{ \textbf{Continuum mechanics simulation of twisted double bilayer graphene.}
		\textbf{(a)} FEA modelling of tDBG unit cell. Based on the atomic structure of tDBG, the ABAB and ABCA are dominant inside a commensurate moiré superlattice; therefore, a rhombic unit cell with periodic boundary conditions is employed to simulate the relaxed tDBG moiré superlattice, where the ABAB domain has a homogeneous eigenstrain of $\varepsilon _{ij}^{\ast }$.
		\textbf{(b)} Three independent strain ($\varepsilon _{xx}$, $\varepsilon _{xy}$ and $\varepsilon _{yy}$) fields from FEA simulation.
		\textbf{(c)} Distributions of four independent strain gradient components ($\varepsilon _{xx,x}$, $\varepsilon _{xx,y}$, $\varepsilon _{yy,x}$ and $\varepsilon _{yy,y}$), which are found to concentrate at domain walls.
		\textbf{(d)} In-plane flexoelectric polarization along $x$- ($P_x$) and $y$-axis ($P_y$). Three types of interfaces emerge, termed as $0^{\circ}$-, $120^{\circ}$- and $240^{\circ}$-domain walls (DWs in figure).
		\textbf{(e)} Total in-plane flexoelectric polarization ($P_{xy}$) from vector summation. ABAB is surrounded by counter clockwise polarizations, while ABCA is encircled by clockwise polarizations. Color denotes the direction of polarization. Three arrows in the insert denote three polarization directions ($0^{\circ}$, $120^{\circ}$ and $240^{\circ}$).
	}
	\label{fig:2} 
\end{figure*}
\begin{figure*}[htbp]	  
	\centering
	\includegraphics[width=6in] {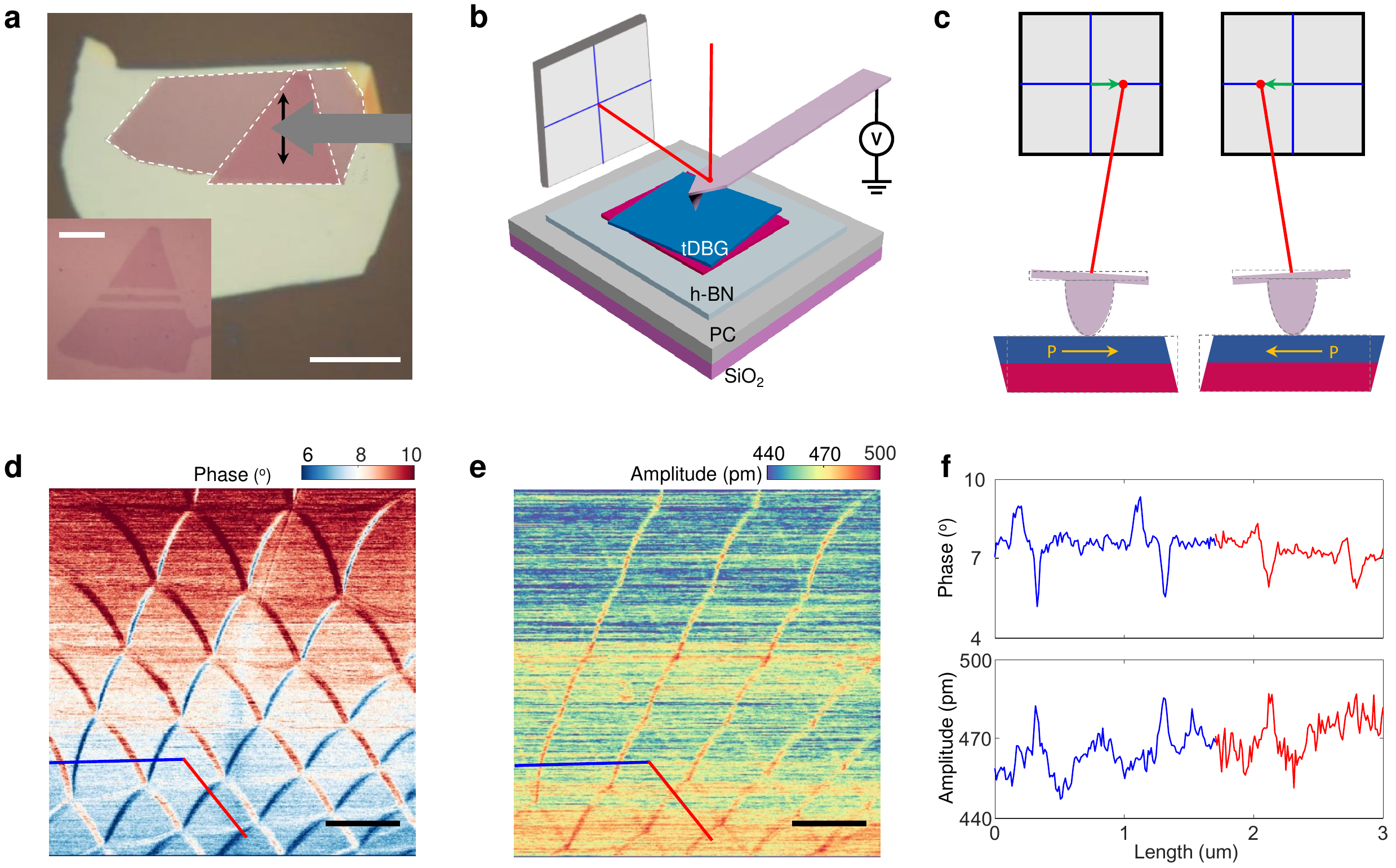}
	\caption
	{ \textbf{Lateral piezoelectric force microscopy mapping in twisted double bilayer graphene.}
		\textbf{(a)} Optical microscope image of tDBG. The light green color in the image represents h-BN, which is at the bottom to support the tDBG; the two pieces of bilayer graphene are on the top. In order to expose the tDBG surface, the sample is flipped over after transfer. The insert shares the optical image of pre-cut bilayer graphene before transfer. Scale bars, 10 $\mu$m.
		\textbf{(b)} Schematic of LPFM setup and stacking order of sample.
		\textbf{(c)} Principle of LPFM.
		\textbf{(d,e)} Phase and amplitude maps of LPFM. The direction of cantilever is parallel to the horizontal. Scale bars, 400 nm.
		\textbf{(f)} Profiles of phase and amplitude for the marked blue and red lines in (d) and (e), respectively. The phase shows only small variation and the amplitude also exhibits only slight change from 450 to 490 pm across the domain walls.
	}
	\label{fig:3} 
\end{figure*}

\begin{figure*}[htbp]	  
	\centering
	\includegraphics[width=6in] {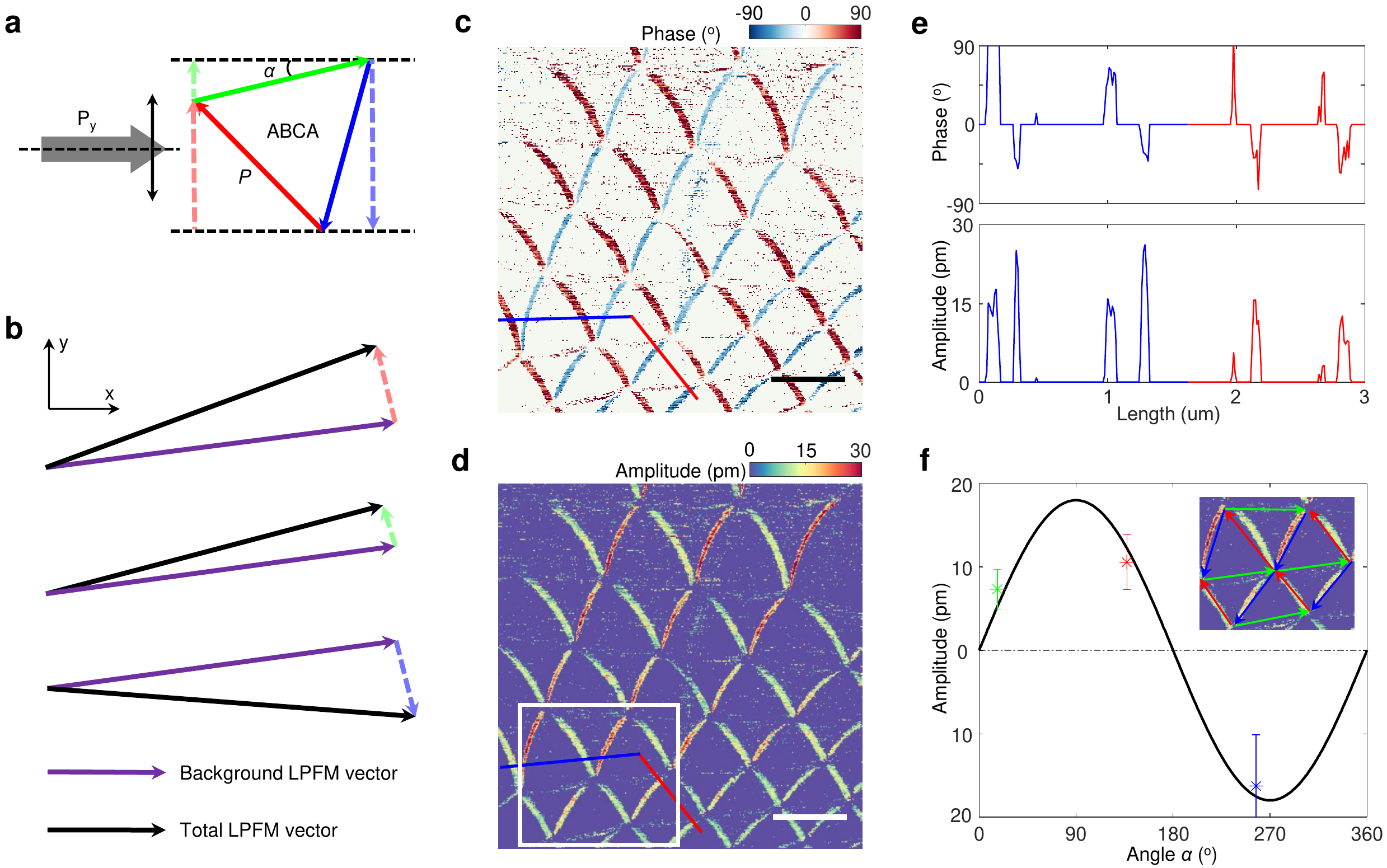}
	\caption
	{ \textbf{Extraction of lateral piezoelectric force microscopy vector.}
		\textbf{(a)} Decomposition of moiré superlattice induced in-plane polarizations surrounding ABCA domain in a single LPFM scan. LPFM can only capture the in-plane polarization components which are perpendicular to the cantilever. For the head-to-end equilateral triangle (solid green, red and blue arrows) polarization vectors, the captured components in a single scan are denoted by dotted arrows.
		\textbf{(b)} Principle of decoupling LPFM signal from the large background. By subtracting a constant background vector, the pure moiré superlattice induced LPFM vector (amplitude and phase) is determined.
		\textbf{(c,d)} Intrinsic moiré superlattice induced LPFM phase and amplitude. The direction of cantilever is illustrated in (a). Scale bars, 400 nm.
		\textbf{(e)} Profiles of phase and amplitude for marked blue and red lines in (c) and (d), wherein domain walls show almost 0 and 180 degree phase contrast.
		\textbf{(f)} Relation between decoupled LPFM amplitude and in-plane polarization angle ($\alpha$). Mean and variance value of decoupled amplitudes for $0^{\circ}$- (green), $120^{\circ}$- (red) and $240^{\circ}$-domain walls (blue) marked in white rectangle of (d) (also see the insert) follow a sinusoidal relation, which matches theoretical predication well.
	}
	\label{fig:4} 
\end{figure*}

\begin{figure*}[htbp]	  
	\centering
	\includegraphics[width=6in] {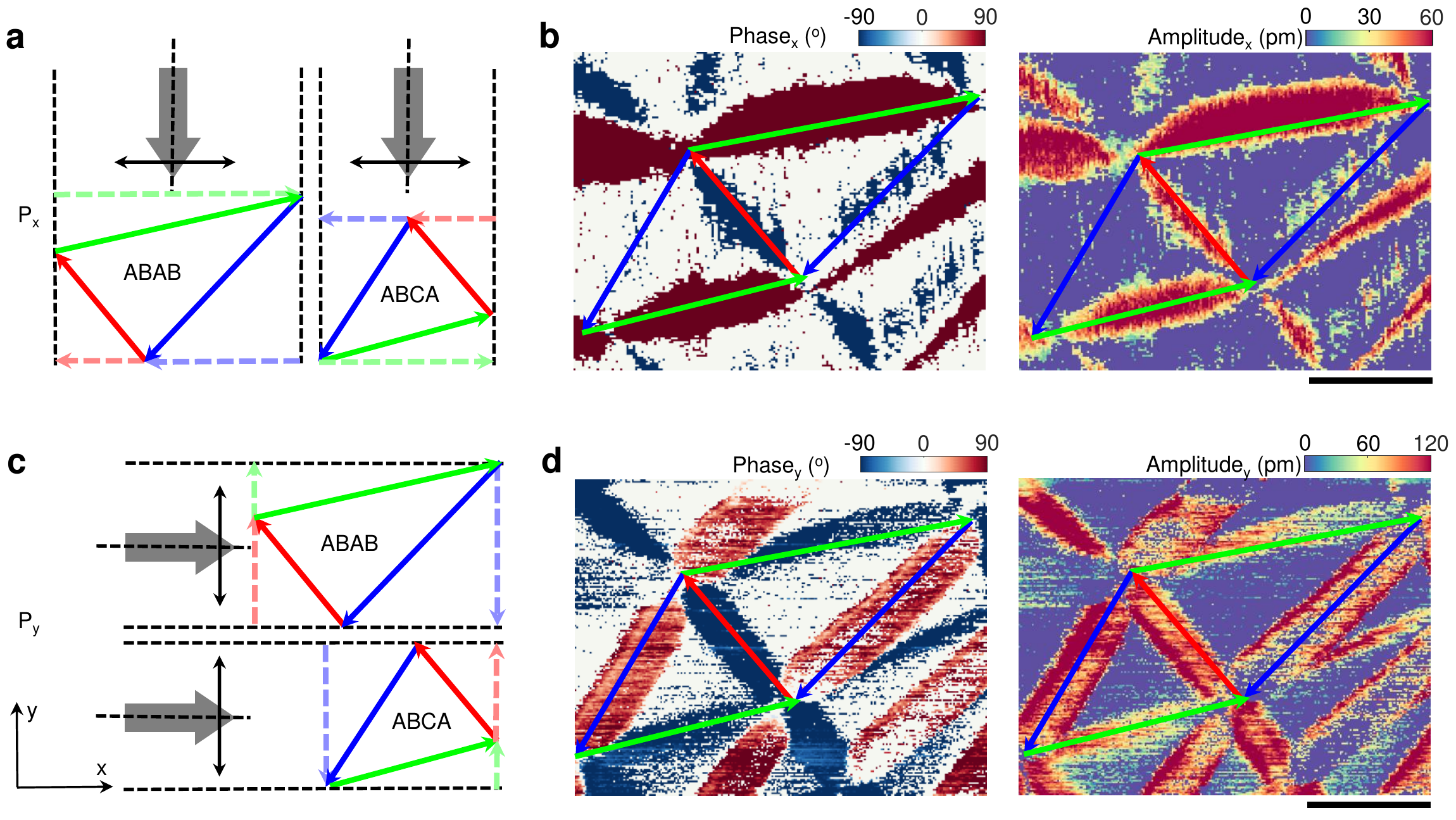}
	\caption
	{ \textbf{Distribution of flexoelectric polarization in twisted double bilayer graphene.}
		\textbf{(a)} Orthogonal decomposition of predicated in-plane polarizations along x-axis around the ABAB and ABCA domains, wherein polarization components of $120^{\circ}$- (red) and $240^{\circ}$-domain walls (blue) are negative, while that of $0^{\circ}$-domain wall is positive.
		\textbf{(b)} Decoupled LPFM phase and amplitude of tDBG in $x$-direction that match well with predictions shown in (a). The direction of cantilever is illustrated in (a). Scale bar, 100 nm.
		\textbf{(c)} Orthogonal decomposition of predicted in-plane polarization along y-axis around ABAB and ABCA domains, wherein the polarization components of $0^{\circ}$- (green) and $120^{\circ}$-domain wall (red) are positive, while that of $240^{\circ}$-domain wall (blue) is negative.
		\textbf{(d)} Decoupled LPFM amplitude and phase of tDBG in y-direction that match well with predictions shown in (c). The direction of cantilever is illustrated in (c). Scale bar, 100 nm.
	}
	\label{fig:5} 
\end{figure*}
\begin{figure}[htbp]	  
	\centering
	\includegraphics[width=3.3in] {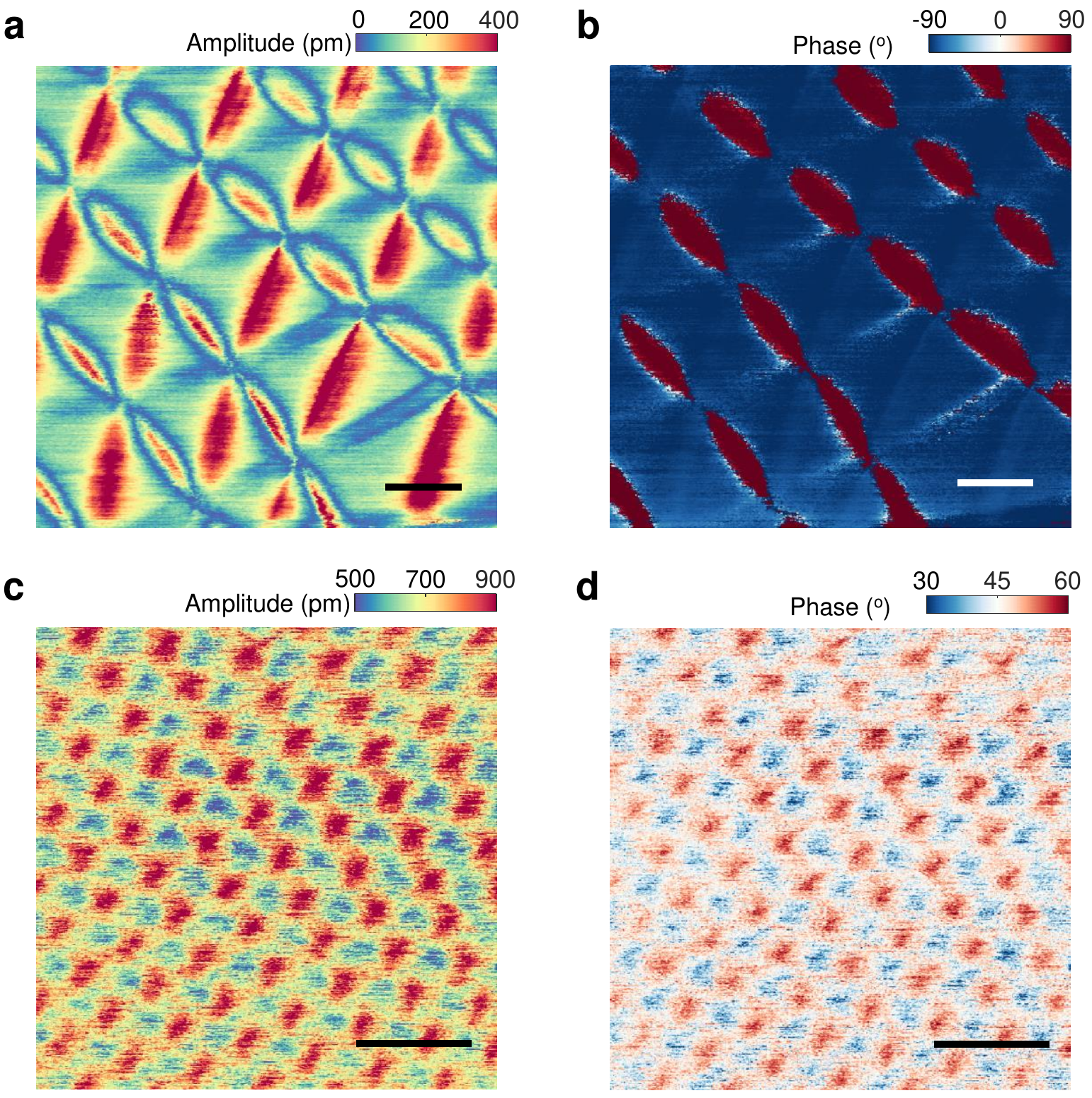}
	\caption
	{ \textbf{Commensurate and incommensurate domains in twisted double bilayer graphene.}
		\textbf{(a,b)} Commensurate domains with well-developed domain walls, wherein twist angle of $\sim$$0.1^{\circ}$ is estimated. The direction of cantilever is parallel to the horizontal. Scale bars, 100 nm.
		\textbf{(c,d)} Incommensurate domains with under-developed domain walls, wherein a twist angle of $\sim$$0.35^{\circ}$ is estimated. The direction of cantilever is parallel to the horizontal. Scale bars, 50 nm.
	}
	\label{fig:6} 
\end{figure}

\section*{Results and Discussions}
We consider two Bernal-stacked (AB) graphene bilayers stacked together; and three different stacking orders may exist as shown in \textbf{Figure 1(a)}, including ABAB, ABBC and ABCA. These domains all belong to $P\overline{3}m1$ space group, and hence are center symmetric, possessing no spontaneous polarization. Furthermore, electronic structure calculations suggest that both ABAB and ABCA domains possess no band gap, as seen in Figure S1(a), in good agreement with a previous report\cite{RN1701}. When a crystal is subjected to another crystal, it can adjust itself to follow the periodicity of the potential under certain conditions, resulting in a commensurate state and the corresponding commensurate domains\cite{RN1684}. Here, two graphene bilayers are twisted with respect to each other by a small in-plane angle $\theta$, and three types of domains emerge in tDBG as shown in \textbf{Figure 1(b)}, providing an ideal platform to study multifarious stacking orders simultaneously. Because ABBC is the most energetically disfavored stacking order\cite{RN1679, RN1587}, ABAB and ABCA expand at the expense of ABBC, forming two large commensurate domains separated by ‘saddle point’ stacking (A-SP-B) interfaces. Such commensurate moiré superlattices dominate in tDBG with small twist angle, and its wavelength can be characterized by $L=(a/2)\cdot \text{csc}(\theta/2)$, where $a$ is the lattice constant of graphene\cite{RN1520}. To better understand this moiré structure, the PES searching\cite{RN1746} between two AB graphene bilayers is calculated, revealing six-fold symmetry as seen in \textbf{Figure 1(c)}. The energetic difference between ABAB and ABCA is negligible, while the line profile indicates an energy barrier between these two domains, corresponding to higher energy at domain walls. Because of the competition between ABAB and ABCA domains (ABAB stacking is more energetically favorable), this six-fold symmetry cannot survive at tiny twist angles\cite{RN1700, RN1697, RN1687, RN1858}. By optimizing a four-layer tDBG moiré superlattice with a $6.01^{\circ}$ twist angle as shown in Figure S1(b), average out-of-plane displacement ($\overline{u}_z$) field in the reconstructed tDBG moiré superlattice is found to be three-fold symmetric (\textbf{Figure 1(d)}). More details and out-of-plane displacement ($u_z$) fields for every layer can be found in Figure S1(c). Notice that the twist angle used in our DFT calculation is larger than experimental value to reduce the computational cost, as the purpose of our DFT calculation is to demonstrate the incompatibility induced by reconstruction, regardless of twist angles. This reconstruction can be understood as the consequence of competition between ABAB and ABCA stacking orders resulting in expansion of ABAB domains, and similar results have been reported in previous studies, showing a convex domain for ABAB and a concave domain for ABCA\cite{RN1687, RN1858}. Furthermore, the lattice mismatch is expected between ABAB and ABCA domains and concentrated across domain walls, leading to inhomogeneous strain distribution. The question arises what the implication of such inhomogeneous strains is.

Since full scale DFT calculations for such moiré superlattices are quite expensive, we turn to coarse-grained continuum analysis, wherein the lattice mismatch between ABAB and ABCA can be represented by an inelastic eigenstrain\cite{RN1687, RN1699}, making it possible to study the system at much larger scale. From the tDBG moiré superlattices shown in \textbf{Figure 1(b)}, a 2D rhombic unit cell representing commensurate ABAB and ABCA domains is adopted as shown in \textbf{Figure 2(a)}, with ABAB domain imposed with an eigenstrain of $\varepsilon _{ij}^{\ast }$ to reflect its lattice mismatch with respect to the other. The stress ($\sigma_{ij}$) and strain ($\varepsilon _{ij}$) distributions that resulting from such eigenstrain are governed by the constitutive equation $\sigma_{ij}=c_{ijkl}\cdot(\varepsilon _{kl}-\varepsilon _{kl}^{\ast })$, wherein $c_{ijkl}$ is the stiffness tensor, and they can be computed by FEA with periodic boundary conditions, as shown in \textbf{Figure 2(b)} and detailed in \textbf{Methods}. Note that we adopt plane-stress condition here in our analysis, which is appropriate for 2D systems while capable of accounting for out-of-plane displacement calculated by DFT. Substantial variations in 2D strain components are observed across domain walls, giving four independent strain gradients as presented in \textbf{Figure 2(c)}, which are clearly concentrated on domain walls. Additional components of strain and strain gradient can be found in Figure S2 and S3, respectively.

The strain gradients at domain walls then lead to flexoelectric polarization\cite{RN1574, RN1575}, even though ABAB and ABCA domains ($P\overline{3}m1$) themselves possess no spontaneous polarization. While graphene is conductive, the electric conductivity and polarization does not necessarily exclude each other, as originally proposed by Anderson and Blount under the concept of polar metal\cite{RN2162} and recently demonstrated experimentally by Shi\cite{RN2163} and Fei\cite{RN1034}. Such flexoelectric polarization ($P_l$) can be evaluated from strain gradients ($\varepsilon _{ij,k}$) as , where $f_{ijkl}$ is the flexoelectric coefficient. Adopting flexoelectric constitutive equations for point groups (3, $\overline{3}$)\cite{RN1677, RN1676}, the in-plane polarizations $P_x$ and $P_y$ in tDBG are calculated as shown in \textbf{Figure 2(d)}, with their vectorial sum presented in \textbf{Figure 2(e)}, revealing $0^{\circ}$-, $120^{\circ}$- and $240^{\circ}$-domain walls consistent with the three-fold symmetric displacement field observed in DFT calculations. More details about FEA can be found in \textbf{Methods}. Note that ABAB domains are surrounded by counter-clockwise polarizations along the domain walls, while the ABCA domains are encircled by domain walls with clockwise polarizations. Such in-plane polarizations in turn induce electromechanical response\cite{RN15, RN1661} that can be measured by LPFM.

To verify the theoretical analysis, tDBG samples are fabricated via a standard ‘tear and stack’ technique with near zero twist angle using a polydimethylsiloxane (PDMS) stamp with polycarbonate (PC) film on the top\cite{RN1686, RN1657}. The optical image of a stacked tDBG sample is presented in \textbf{Figure 3(a)}. The inset shows the bilayer graphene before transferring, ensuring the same lattice orientation before twisting. Details about the fabrication can be found in \textbf{Methods} and Figure S4. LPFM is then employed to measure the in-plane electromechanical response of the tDBG at the nanoscale, as schematically shown in \textbf{Figure 3(b)}. The applied voltage at the probe produces a vertical electric field under the tip, which excites in-plane shear deformation of the material when there are in-plane polarizations\cite{RN110, RN15}. Note that our tDBG sample is only around 1.4 nm thick, which is below the typical screening length scale\cite{RN1034}, and thus it can be sufficiently penetrated by vertical electric field for PFM measurement despite good in-plane conductivity of the graphene. Importantly, in-plane piezoresponse only occurs when the polarization is perpendicular to the cantilever axis, so opposite polarizations induce piezoresponse with $180^{\circ}$ phase contrast, as schematically shown in \textbf{Figure 3(c)}.

Maps of experimentally measured LPFM phase and amplitude in \textbf{Figure 3(d-e)} reveal the anticipated moiré pattern of tDBG with an estimated twist angle of $\theta \sim$$0.035^{\circ}$ from the moiré superlattice length $L$ of approximately 400 nm. It is observed that domain walls exhibit significant variations in contrast depending on their orientations, whereas there is negligible contrast difference among domains. Vertical contrast in the maps is induced by a constant shift among every scan line, which can be eliminated by flattening, as shown in Figure S5. LPFM scans over a larger area in Figure S6 exhibit similar moiré patterns. The shapes of domains are also informative, with ABAB domains convex and ABCA domains concave, consistent with both previous works\cite{RN1687, RN1858} and our FEA simulations.

However, there are two unexpected observations in the phase and amplitude line profiles along the blue and red lines marked in \textbf{Figure 3(d-e)}. First, there is a phase contrast of just $4^{\circ}$, rather than expected $180^{\circ}$ between opposite polarizations. Second, there is a non-zero amplitude inside center symmetric ABAB and ABCA domains. The measured amplitudes are around 465 pm inside the domains and the differences between domains and domain walls are only about 15 pm, as shown in \textbf{Figure 3(f)}, suggesting that LPFM amplitudes are dominated by the signal inside domains. Similar phenomena have also been observed previously\cite{RN1679}, and the question is how we can reconcile the inconsistency between theoretical expectations and experimental observations. Furthermore, there exist out-of-plane electromechanical responses as probed by vertical PFM (VPFM), though they are much weaker than in-plane ones as compared in Figure S7. They could arise from in-plane response via Poisson’s effect\cite{RN2064}.

PFM signals are quite complex, arising from the interplay between piezoelectricity, electrostatic interaction, electrochemical strain and even Joule heating\cite{RN115}. These mechanisms may contribute as a background PFM signal on top of a relatively small flexoelectric response\cite{RN1662}. To extract the intrinsic flexoelectric contribution via vectorial analysis\cite{RN1689}, we examine three domain walls surrounding an ABCA domain as shown in \textbf{Figure 4(a)}. LPFM captures the lateral signal of the cantilever, and thus can only reflect polarization components that are perpendicular to the cantilever. As a result, for the head to end clock-wise polarizations around an ABCA domain with a small tilt angle $\alpha $ between $0^{\circ}$-domain wall and cantilever, the magnitudes of polarizations at three domain walls captured by LPFM are $P\cdot \sin \alpha $, $P\cdot \sin{\alpha+120^{\circ}} $ and $P\cdot \sin{\alpha+240^{\circ}} $, respectively, while the phase contrast is either $0^{\circ}$ or $180^{\circ}$. However, when there is a large constant background imposed on top of the flexoelectric signal, then vectorial summation as shown in \textbf{Figure 4(b)} results in the total LPFM response having only small differences in amplitude and phase at these domain walls, which is what we have observed experimentally in \textbf{Figures 3(d-e)}.

With the above understanding we can now remove the background signal from the LPFM data, recovering the LPFM phase and amplitude of tDBG shown in \textbf{Figures 4(c-d)} corresponding to the flexoelectric polarizations only. The decoupling process is explained in Figure S8. Here $P_y$ is measured by the horizontal cantilever as shown in \textbf{Figure 4(a)}. Both decoupled phase and amplitude maps agree well with theoretically predicted $P_y$ in \textbf{Figure 2(d)}, wherein the $0^{\circ}$-domain wall is invisible and the other two exhibit almost $180^{\circ}$ phase contrast and similar amplitudes, as evident from the line profiles in \textbf{Figure 4(e)}. By summing the decoupled amplitudes at domain walls inside the white square of \textbf{Figure 4(d)} (also see the insert in \textbf{Figure 4(f)}), the mean and variance of $0^{\circ}$- (green arrows), $120^{\circ}$- (red arrows) and $240^{\circ}$-domain walls (blue arrows), shown in \textbf{Figure 4(f)}, are fully consistent with the expected sinusoidal variation. These data thus confirm that after removing a large background signal via vectorial analysis, we have recovered the intrinsic flexoelectric response at domain walls of tDBG.

To capture all in-plane polarization components, PFM measurements at two different angles (0 and 90 degrees) between the cantilever and sample are necessary. The intrinsic flexoelectric response can then be extracted via the vectorial analysis, as discussed above. To this end, \textbf{Figure 5(a)} gives the resolved polarization components around ABAB and ABCA domains in the x axis direction ($P_x$), with that of $120^{\circ}$- (red dotted arrows) and $240^{\circ}$-domain walls (blue dotted arrows) being negative, while that of $0^{\circ}$-domain walls (green dotted arrows) being positive. The decoupled LPFM phase and amplitude are then shown in \textbf{Figure 5(b)}, reconstructed from raw experimental data in Figure S9. As expected, $0^{\circ}$-domain walls (paralleling green arrows) has near $90^{\circ}$ phase, while $120^{\circ}$- (paralleling red arrows) and $240^{\circ}$-domain walls (paralleling blue arrows) have -$90^{\circ}$ phase. The largest amplitude is also seen for $0^{\circ}$-domain walls. Rotating the cantilever by $\sim$$90^{\circ}$, $P_y$ is measured as shown in textbf{Figure 5(c)}. The corresponding decoupled LPFM maps are shown in \textbf{Figure 5(d)}, matching well with the predicted results. Here, $0^{\circ}$- (paralleling green arrows) and $120^{\circ}$-domain walls (paralleling red arrows) exhibit positive polarization while $240^{\circ}$-domain walls (paralleling blue arrows) have negative polarization, as revealed by the corresponding phase contrast. One point worth noting is that it is difficult to determine the absolute direction of in-plane polarization via LPFM, yet we can get the guidance on the polarization direction from our FEA simulation.

For a tDBG with small twist angle, its moiré superlattices can adjust themselves into a commensurate state, forming large commensurate ABAB and ABCA domains separated by A-SP-B domain walls. When the twist angle increases above a critical value, the superlattices of tDBG hold without forming large commensurate domains, leading to incommensurate state\cite{RN1684} and the corresponding incommensurate domains. It turns out that both commensurate and incommensurate domains of tDBG can be captured by LPFM, as presented in \textbf{Figure 6}. For commensurate domains in \textbf{Figure 6(a-b)}, the amplitude map shows clearly three well-developed domain walls and the corresponding phase contrast is close to $180^{\circ}$, suggesting opposite polarization directions. The length of the moiré pattern ranges from 100 to 200 nm, corresponding to $0.14^{\circ}$ to $0.07^{\circ}$ twist angles. When the twist angle increases to $\sim$$0.35^{\circ}$, the moiré pattern length drops to 40 nm, resulting in incommensurate domains as seen in \textbf{Figure 6(c-d)}. Here domain walls are under-developed and the lattice mismatch, as well as flexoelectricity, are concentrated in ABBC stacking regions with comparable size to ABAB and ABCA domains. This finding is consistent with previous results\cite{RN1697} as well as our DFT results in \textbf{Figure 1(c)}.

\section*{Conclusions}
While moiré superlattices such as tBG and tDBG are thought to exhibit appreciable flexoelectric effect, only unexpectedly small phase contrast was observed in tBG via PFM so far, demanding a more complete theoretical understanding of this emergent phenomena. To identify flexoelectricity in moiré superlattices, we have evaluated the strain gradient and the corresponding flexoelectric polarization using coarse-grain DFT calculations and FEA simulations. We found that the in-plane polarization is three-fold symmetric at domain walls, counter-clockwise around ABAB and clockwise around ABCA. These predictions have been confirmed experimentally by LPFM, wherein the intrinsic flexoelectric response can be isolated from background by vectorial analysis, yielding distributions of LPFM phase and amplitude at domain walls that are fully consistent with our theoretical predications. Our work provides a general methodology for studying flexoelectricity in tDBG and beyond while providing new insights into symmetry engineered breaking of centrosymmetric materials.

\section*{Methods}
\noindent\textbf{Density functional theory.} First-principles calculations using DFT implemented in the Vienna Ab-Initial Package (VASP) were performed\cite{RN1690, RN1274}. The electron and core interactions are included using the frozen-core projected augmented wave (PAW) approach\cite{RN1271, RN1275}, and the generalized gradient approximation (GGA) formulated by Perdew, Burke, and Ernzerhof (PBE) was adopted\cite{RN1276}. The van der Walls correction was accounted with the Grimme DFT-D3 functional\cite{RN1691}. All related structures were optimized by the recommended conjugate-gradient algorithm until the maximum atomic force component acting on each atom is less than 0.01 eV/Å. To calculate the band structure of ABAB or ABCA stacking graphene, the primitive cell with periodic length of 2.47 Å was modeled and the vacuum distance larger than 15 Å was used to avoid the interference of adjacent images along the $z$ direction. Using the same primitive cell, the potential energy was searched using a step size of 0.02 Å in the $x$ and $y$ directions of all the atomic coordinates, while the z axis was fixed. Finally, a tDBG supercell with a twist angle of $6.01^{\circ}$ (moiré superlattice length is 23.56 Å) was constructed to simulate the lattice optimization, as shown in Figure S1(b), which comprises of 728 carbon atoms.
\\ \hspace*{\fill} \\
\noindent\textbf{Finite element analysis.} COMSOL Multiphysics software was chosen to simulate this problem because of its parametric modeling and convenient definition of initial strain. The commensurate ABAB and ABCA domains were setup in COMSOL to simulate the relaxation of two domains. Because only the eigenstrain difference between ABAB and ABCA domains can affect the deformation of domains, the eigenstrain in ABCA domain was set to zero and the eigenstrain of ABAB domain was set to 0.5 to magnify their difference. After calculating their displacement fields, their strain and strain gradient fields were deduced from displacement fields of ABAB and ABCA domains, as shown in Figure S2 and S3. Furthermore, the polarizations were calculated based on the flexoelectric constitutive equations shown above. For point groups (3, $\overline{3}$), the flexoelectric coefficient ($f_{ijkl}$) has the following form
\begin{widetext}
\begin{equation}
f_{ijkl}=\begin{pmatrix}
	f_{1111} &f_{1112}  & (f_{1111}-f_{1122}) & -f_{2111}  & f_{1121} & f_{1122} \\ 
	f_{2111} & (f_{1111}-f_{1122}) & -f_{1112} & f_{1111} & (f_{1111}-f_{1122}) & -f_{1121}
\end{pmatrix}
\end{equation}
\end{widetext} And the value is set as
\begin{equation}
	f_{ijkl}=\begin{pmatrix}
		0 &1  & 0 & 1  & 0 & 0 \\ 
		-1 & 0 & -1 & 0 & 0 & 0
	\end{pmatrix}
\end{equation}in our computation. The responsible strain gradient for the observed response are $\varepsilon _{xx,x}$, $\varepsilon _{xx,y}$, $\varepsilon _{yy,x}$ and $\varepsilon _{yy,y}$. Substituting flexoelectric coefficient ($f_{ijkl}$) and strain gradients ($\varepsilon _{ij,k}$) into flexoelectric constitutive equations, the flexoelectric polarizations are determined.
\\ \hspace*{\fill} \\
\noindent\textbf{Sample fabrication.} Our fabrication of tDBG samples followed a standard ‘tear and stack’ technique, with first cutting the bilayer graphene by an atomic force microscope tip to relieve its strain during assembly. A PC film on top of PDMS stamp was used to pick up h-BN, then half of bilayer graphene, followed by the second half of graphene with a desired twist angle. To access the surface of tDBG, the PC film was released from stamp by the thermal release tape, and then flipped over and placed on a Si/SiO$_2$ chip, as shown in Figure S4.
\\ \hspace*{\fill} \\
\noindent\textbf{Piezoelectric force microscopy.} LPFM experiments (\textbf{Figure 3} and Figure S6) were performed on a Bruker Dimension Icon in ambient using SCM-PIT-V2 probe with spring constant $\sim$3 N/m, free resonance frequency $\sim$60 kHz and LPFM resonance frequency $\sim$720 kHz. By tuning the single excitation frequency, the lateral piezoelectric signals of tDBG can be stably captured by the single frequency technique without significant shift because of the atomic flat surface of tDBG. Other LPFM experiments were conducted on an Oxford Instruments Asylum Research MFP-3D Infinity atomic force microscope. ASYELEC-01 Ti/Ir coated silicon probes with a spring constant $\sim$3 N/m and free resonance frequency $\sim$75 kHz were used to carry out single frequency measurement. Using the coordinate system as shown in \textbf{Figure 2}, all cantilevers in LPFM measurement are parallel to $x$ axis except results in \textbf{Figure 5(b)} and its related raw results as shown in Figure S9(a-c) which is parallel to $y$ axis. VPFM experiments were also conducted on an Oxford Instruments Asylum Research MFP-3D Infinity atomic force microscope. ASYELEC-01 Ti/Ir coated silicon probes with a spring constant $\sim$3 N/m and free resonance frequency $\sim$75 kHz were used to carry out single frequency measurement.

\section*{Acknowledgments}
This project is mainly supported by the U.S. NSF through the UW Molecular Engineering Materials Center, a Materials Research Science and Engineering Center (DMR-1719797). PFM measurements are partially supported by Programmable Quantum Materials, an Energy Frontier Research Center funded by the U.S. Department of Energy (DOE), Office of Science, Basic Energy Sciences (BES), under award DE-SC0019443. X.X. acknowledges support from the Boeing Distinguished Professorship in Physics. X.X. and M.Y. acknowledge support from the State of Washington funded Clean Energy Institute. K.W. and T.T. were supported by the Elemental Strategy Initiative conducted by the MEXT, Japan, and the CREST (JPMJCR15F3), JST. J. L. acknowledges the support from National Key Research and Development Program of China (2016YFA0201001), Shenzhen Science and Technology Innovation Committee (JCYJ20170818163902553), the Leading Talents Program of Guangdong Province (2016LJ06C372), and the Guangdong Provincial Key Laboratory Program from the Department of Science and Technology of Guangdong Province (2021B1212040001). Xiao W. acknowledges National Natural Science Foundation of China (22003074) and supports by Center for Computational Science and Engineering at Southern University of Science and Technology. Y. L. also thanks the support of the China Scholarship Council.

\section*{Author contributions}
X.X. and Y.L. conceived the experiment, and J.L. supervised the data analysis and computations. Y.L. fabricated the devices. Y.L. performed the measurement assisted by Xi W. Xiao W. and D.T. carried out the DFT calculation. Y.L. performed the FEA simulation. K.W. and T.T. provided the bulk BN crystals. Y.L. and J.L. analyzed the data and wrote the paper with input from X.X., M.Y., D.C., and D.G. All authors discussed the results.

\section*{Competing interests}
The authors declare no competing interests.

\section*{Data availability}
The data that support the plots within this paper and other findings of this study are available from the corresponding author upon reasonable request.

\bibliographystyle{naturemag}
\bibliography{references_flexoelectricity}

\begin{thebibliography}{10}
\expandafter\ifx\csname url\endcsname\relax
  \def\url#1{\texttt{#1}}\fi
\expandafter\ifx\csname urlprefix\endcsname\relax\def\urlprefix{URL }\fi
\providecommand{\bibinfo}[2]{#2}
\providecommand{\eprint}[2][]{\url{#2}}

\bibitem{RN1603}
\bibinfo{author}{Balents, L.}, \bibinfo{author}{Dean, C.~R.},
  \bibinfo{author}{Efetov, D.~K.} \& \bibinfo{author}{Young, A.~F.}
\newblock \bibinfo{title}{Superconductivity and strong correlations in moiré
  flat bands}.
\newblock \emph{\bibinfo{journal}{Nat. Phys.}} \textbf{\bibinfo{volume}{16}},
  \bibinfo{pages}{725–733} (\bibinfo{year}{2020}).

\bibitem{RN989}
\bibinfo{author}{Cao, Y.} \emph{et~al.}
\newblock \bibinfo{title}{Unconventional superconductivity in magic-angle
  graphene superlattices}.
\newblock \emph{\bibinfo{journal}{Nature}} \textbf{\bibinfo{volume}{556}},
  \bibinfo{pages}{43--50} (\bibinfo{year}{2018}).
\newblock \urlprefix\url{https://www.ncbi.nlm.nih.gov/pubmed/29512651}.

\bibitem{RN1515}
\bibinfo{author}{Sharpe, A.~L.} \emph{et~al.}
\newblock \bibinfo{title}{Emergent ferromagnetism near three-quarters filling
  in twisted bilayer graphene}.
\newblock \emph{\bibinfo{journal}{Science}} \textbf{\bibinfo{volume}{365}},
  \bibinfo{pages}{605--608} (\bibinfo{year}{2019}).
\newblock
  \urlprefix\url{https://science.sciencemag.org/content/sci/365/6453/605.full.pdf}.

\bibitem{RN1780}
\bibinfo{author}{Serlin, M.} \emph{et~al.}
\newblock \bibinfo{title}{Intrinsic quantized anomalous hall effect in a moiré
  heterostructure}.
\newblock \emph{\bibinfo{journal}{Science}} \textbf{\bibinfo{volume}{367}},
  \bibinfo{pages}{900--903} (\bibinfo{year}{2020}).
\newblock
  \urlprefix\url{https://science.sciencemag.org/content/sci/367/6480/900.full.pdf}.

\bibitem{RN1785}
\bibinfo{author}{Zheng, Z.} \emph{et~al.}
\newblock \bibinfo{title}{Unconventional ferroelectricity in moire
  heterostructures}.
\newblock \emph{\bibinfo{journal}{Nature}} \textbf{\bibinfo{volume}{588}},
  \bibinfo{pages}{71--76} (\bibinfo{year}{2020}).
\newblock \urlprefix\url{https://www.ncbi.nlm.nih.gov/pubmed/33230334}.

\bibitem{RN1679}
\bibinfo{author}{McGilly, L.~J.} \emph{et~al.}
\newblock \bibinfo{title}{Visualization of moire superlattices}.
\newblock \emph{\bibinfo{journal}{Nat. Nanotechnol.}}
  \textbf{\bibinfo{volume}{15}}, \bibinfo{pages}{580–584}
  (\bibinfo{year}{2020}).
\newblock \urlprefix\url{https://www.ncbi.nlm.nih.gov/pubmed/32572229}.

\bibitem{RN1721}
\bibinfo{author}{He, M.} \emph{et~al.}
\newblock \bibinfo{title}{Symmetry breaking in twisted double bilayer
  graphene}.
\newblock \emph{\bibinfo{journal}{Nat. Phys.}}  (\bibinfo{year}{2020}).

\bibitem{RN1695}
\bibinfo{author}{Liu, X.} \emph{et~al.}
\newblock \bibinfo{title}{Tunable spin-polarized correlated states in twisted
  double bilayer graphene}.
\newblock \emph{\bibinfo{journal}{Nature}} \textbf{\bibinfo{volume}{583}},
  \bibinfo{pages}{221--225} (\bibinfo{year}{2020}).
\newblock \urlprefix\url{https://www.ncbi.nlm.nih.gov/pubmed/32641816}.

\bibitem{RN1562}
\bibinfo{author}{Shen, C.} \emph{et~al.}
\newblock \bibinfo{title}{Correlated states in twisted double bilayer
  graphene}.
\newblock \emph{\bibinfo{journal}{Nat. Phys.}} \textbf{\bibinfo{volume}{16}},
  \bibinfo{pages}{520--525} (\bibinfo{year}{2020}).

\bibitem{RN2142}
\bibinfo{author}{Cross, L.~E.}
\newblock \bibinfo{title}{Flexoelectric effects: Charge separation in
  insulating solids subjected to elastic strain gradients}.
\newblock \emph{\bibinfo{journal}{Journal of Materials Science}}
  \textbf{\bibinfo{volume}{41}}, \bibinfo{pages}{53--63}
  (\bibinfo{year}{2006}).
\newblock \urlprefix\url{https://doi.org/10.1007/s10853-005-5916-6}.

\bibitem{RN2143}
\bibinfo{author}{Ma, W.} \& \bibinfo{author}{Cross, L.~E.}
\newblock \bibinfo{title}{Large flexoelectric polarization in ceramic lead
  magnesium niobate}.
\newblock \emph{\bibinfo{journal}{Applied Physics Letters}}
  \textbf{\bibinfo{volume}{79}}, \bibinfo{pages}{4420--4422}
  (\bibinfo{year}{2001}).
\newblock \urlprefix\url{https://doi.org/10.1063/1.1426690}.

\bibitem{RN2141}
\bibinfo{author}{Ma, W.} \& \bibinfo{author}{Cross, L.~E.}
\newblock \bibinfo{title}{Observation of the flexoelectric effect in relaxor
  pb(mg1/3nb2/3)o3 ceramics}.
\newblock \emph{\bibinfo{journal}{Applied Physics Letters}}
  \textbf{\bibinfo{volume}{78}}, \bibinfo{pages}{2920--2921}
  (\bibinfo{year}{2001}).
\newblock \urlprefix\url{https://doi.org/10.1063/1.1356444}.

\bibitem{RN1575}
\bibinfo{author}{Wang, B.}, \bibinfo{author}{Gu, Y.}, \bibinfo{author}{Zhang,
  S.} \& \bibinfo{author}{Chen, L.-Q.}
\newblock \bibinfo{title}{Flexoelectricity in solids: Progress, challenges, and
  perspectives}.
\newblock \emph{\bibinfo{journal}{Prog. Mater. Sci.}}
  \textbf{\bibinfo{volume}{106}}, \bibinfo{pages}{100570}
  (\bibinfo{year}{2019}).

\bibitem{RN1773}
\bibinfo{author}{Nataly~Chen, Q.} \emph{et~al.}
\newblock \bibinfo{title}{Delineating local electromigration for nanoscale
  probing of lithium ion intercalation and extraction by electrochemical strain
  microscopy}.
\newblock \emph{\bibinfo{journal}{Appl. Phys. Lett.}}
  \textbf{\bibinfo{volume}{101}} (\bibinfo{year}{2012}).

\bibitem{RN137}
\bibinfo{author}{Vasudevan, R.~K.}, \bibinfo{author}{Balke, N.},
  \bibinfo{author}{Maksymovych, P.}, \bibinfo{author}{Jesse, S.} \&
  \bibinfo{author}{Kalinin, S.~V.}
\newblock \bibinfo{title}{Ferroelectric or non-ferroelectric: Why so many
  materials exhibit “ferroelectricity” on the nanoscale}.
\newblock \emph{\bibinfo{journal}{Appl. Phys. Rev.}}
  \textbf{\bibinfo{volume}{4}}, \bibinfo{pages}{021302} (\bibinfo{year}{2017}).

\bibitem{RN2164}
\bibinfo{author}{Vizner~Stern, M.} \emph{et~al.}
\newblock \bibinfo{title}{Interfacial ferroelectricity by van der waals
  sliding}.
\newblock \emph{\bibinfo{journal}{Science}} \bibinfo{pages}{eabe8177}
  (\bibinfo{year}{2021}).
\newblock
  \urlprefix\url{https://science.sciencemag.org/content/sci/early/2021/06/09/science.abe8177.full.pdf}.

\bibitem{RN2145}
\bibinfo{author}{Woods, C.~R.} \emph{et~al.}
\newblock \bibinfo{title}{Charge-polarized interfacial superlattices in
  marginally twisted hexagonal boron nitride}.
\newblock \emph{\bibinfo{journal}{Nature Communications}}
  \textbf{\bibinfo{volume}{12}}, \bibinfo{pages}{347} (\bibinfo{year}{2021}).
\newblock \urlprefix\url{https://doi.org/10.1038/s41467-020-20667-2}.

\bibitem{RN2152}
\bibinfo{author}{Yasuda, K.}, \bibinfo{author}{Wang, X.},
  \bibinfo{author}{Watanabe, K.}, \bibinfo{author}{Taniguchi, T.} \&
  \bibinfo{author}{Jarillo-Herrero, P.}
\newblock \bibinfo{title}{Stacking-engineered ferroelectricity in bilayer boron
  nitride}.
\newblock \emph{\bibinfo{journal}{Science}}  (\bibinfo{year}{2021}).

\bibitem{RN1662}
\bibinfo{author}{Zubko, P.}, \bibinfo{author}{Catalan, G.} \&
  \bibinfo{author}{Tagantsev, A.~K.}
\newblock \bibinfo{title}{Flexoelectric effect in solids}.
\newblock \emph{\bibinfo{journal}{Annu. Rev. Mater. Res.}}
  \textbf{\bibinfo{volume}{43}}, \bibinfo{pages}{387--421}
  (\bibinfo{year}{2013}).

\bibitem{RN1689}
\bibinfo{author}{Jungk, T.}, \bibinfo{author}{Hoffmann, $\widetilde{A}$.} \&
  \bibinfo{author}{Soergel, E.}
\newblock \bibinfo{title}{Quantitative analysis of ferroelectric domain imaging
  with piezoresponse force microscopy}.
\newblock \emph{\bibinfo{journal}{Appl. Phys. Lett.}}
  \textbf{\bibinfo{volume}{89}}, \bibinfo{pages}{163507}
  (\bibinfo{year}{2006}).

\bibitem{RN1701}
\bibinfo{author}{Latil, S.} \& \bibinfo{author}{Henrard, L.}
\newblock \bibinfo{title}{Charge carriers in few-layer graphene films}.
\newblock \emph{\bibinfo{journal}{Phys. Rev. Lett.}}
  \textbf{\bibinfo{volume}{97}}, \bibinfo{pages}{036803}
  (\bibinfo{year}{2006}).
\newblock \urlprefix\url{https://www.ncbi.nlm.nih.gov/pubmed/16907528}.

\bibitem{RN1684}
\bibinfo{author}{Woods, C.~R.} \emph{et~al.}
\newblock \bibinfo{title}{Commensurate–incommensurate transition in graphene
  on hexagonal boron nitride}.
\newblock \emph{\bibinfo{journal}{Nature Physics}}
  \textbf{\bibinfo{volume}{10}}, \bibinfo{pages}{451--456}
  (\bibinfo{year}{2014}).

\bibitem{RN1587}
\bibinfo{author}{Yoo, H.} \emph{et~al.}
\newblock \bibinfo{title}{Atomic and electronic reconstruction at the van der
  waals interface in twisted bilayer graphene}.
\newblock \emph{\bibinfo{journal}{Nat. Mater.}} \textbf{\bibinfo{volume}{18}},
  \bibinfo{pages}{448--453} (\bibinfo{year}{2019}).
\newblock \urlprefix\url{https://www.ncbi.nlm.nih.gov/pubmed/30988451}.

\bibitem{RN1520}
\bibinfo{author}{Burg, G.~W.} \emph{et~al.}
\newblock \bibinfo{title}{Correlated insulating states in twisted double
  bilayer graphene}.
\newblock \emph{\bibinfo{journal}{Phys. Rev. Lett.}}
  \textbf{\bibinfo{volume}{123}}, \bibinfo{pages}{197702}
  (\bibinfo{year}{2019}).
\newblock \urlprefix\url{https://www.ncbi.nlm.nih.gov/pubmed/31765206}.

\bibitem{RN1746}
\bibinfo{author}{Reguzzoni, M.}, \bibinfo{author}{Fasolino, A.},
  \bibinfo{author}{Molinari, E.} \& \bibinfo{author}{Righi, M.~C.}
\newblock \bibinfo{title}{Potential energy surface for graphene on graphene:ab
  initioderivation, analytical description, and microscopic interpretation}.
\newblock \emph{\bibinfo{journal}{Phys. Rev. B}} \textbf{\bibinfo{volume}{86}},
  \bibinfo{pages}{245434} (\bibinfo{year}{2012}).

\bibitem{RN1700}
\bibinfo{author}{Aoki, M.} \& \bibinfo{author}{Amawashi, H.}
\newblock \bibinfo{title}{Dependence of band structures on stacking and field
  in layered graphene}.
\newblock \emph{\bibinfo{journal}{Solid State Commun.}}
  \textbf{\bibinfo{volume}{142}}, \bibinfo{pages}{123--127}
  (\bibinfo{year}{2007}).

\bibitem{RN1697}
\bibinfo{author}{Gargiulo, F.} \& \bibinfo{author}{Yazyev, O.~V.}
\newblock \bibinfo{title}{Structural and electronic transformation in low-angle
  twisted bilayer graphene}.
\newblock \emph{\bibinfo{journal}{2D Materials}} \textbf{\bibinfo{volume}{5}},
  \bibinfo{pages}{015019} (\bibinfo{year}{2017}).

\bibitem{RN1687}
\bibinfo{author}{Hattendorf, S.}, \bibinfo{author}{Georgi, A.},
  \bibinfo{author}{Liebmann, M.} \& \bibinfo{author}{Morgenstern, M.}
\newblock \bibinfo{title}{Networks of aba and abc stacked graphene on mica
  observed by scanning tunneling microscopy}.
\newblock \emph{\bibinfo{journal}{Surf. Sci.}} \textbf{\bibinfo{volume}{610}},
  \bibinfo{pages}{53--58} (\bibinfo{year}{2013}).

\bibitem{RN1858}
\bibinfo{author}{Kerelsky, A.} \emph{et~al.}
\newblock \bibinfo{title}{Moiréless correlations in abca graphene}.
\newblock \emph{\bibinfo{journal}{Proceedings of the National Academy of
  Sciences}} \textbf{\bibinfo{volume}{118}}, \bibinfo{pages}{e2017366118}
  (\bibinfo{year}{2021}).
\newblock
  \urlprefix\url{https://www.pnas.org/content/pnas/118/4/e2017366118.full.pdf}.

\bibitem{RN1699}
\bibinfo{author}{Tirry, W.} \& \bibinfo{author}{Schryvers, D.}
\newblock \bibinfo{title}{Linking a completely three-dimensional nanostrain to
  a structural transformation eigenstrain}.
\newblock \emph{\bibinfo{journal}{Nat. Mater.}} \textbf{\bibinfo{volume}{8}},
  \bibinfo{pages}{752--757} (\bibinfo{year}{2009}).
\newblock \urlprefix\url{https://www.ncbi.nlm.nih.gov/pubmed/19543276}.

\bibitem{RN1574}
\bibinfo{author}{Shen, S.} \& \bibinfo{author}{Hu, S.}
\newblock \bibinfo{title}{A theory of flexoelectricity with surface effect for
  elastic dielectrics}.
\newblock \emph{\bibinfo{journal}{J. Mech. Phys. Solids}}
  \textbf{\bibinfo{volume}{58}}, \bibinfo{pages}{665--677}
  (\bibinfo{year}{2010}).

\bibitem{RN2162}
\bibinfo{author}{Anderson, P.~W.} \& \bibinfo{author}{Blount, E.~I.}
\newblock \bibinfo{title}{Symmetry considerations on martensitic
  transformations: "ferroelectric" metals?}
\newblock \emph{\bibinfo{journal}{Physical Review Letters}}
  \textbf{\bibinfo{volume}{14}}, \bibinfo{pages}{217--219}
  (\bibinfo{year}{1965}).
\newblock \urlprefix\url{https://link.aps.org/doi/10.1103/PhysRevLett.14.217}.

\bibitem{RN2163}
\bibinfo{author}{Shi, Y.} \emph{et~al.}
\newblock \bibinfo{title}{A ferroelectric-like structural transition in a
  metal}.
\newblock \emph{\bibinfo{journal}{Nature Materials}}
  \textbf{\bibinfo{volume}{12}}, \bibinfo{pages}{1024--1027}
  (\bibinfo{year}{2013}).
\newblock \urlprefix\url{https://doi.org/10.1038/nmat3754}.

\bibitem{RN1034}
\bibinfo{author}{Fei, Z.} \emph{et~al.}
\newblock \bibinfo{title}{Ferroelectric switching of a two-dimensional metal}.
\newblock \emph{\bibinfo{journal}{Nature}}  (\bibinfo{year}{2018}).

\bibitem{RN1677}
\bibinfo{author}{Shu, L.} \emph{et~al.}
\newblock \bibinfo{title}{Relationship between direct and converse
  flexoelectric coefficients}.
\newblock \emph{\bibinfo{journal}{J. Appl. Phys.}}
  \textbf{\bibinfo{volume}{116}}, \bibinfo{pages}{144105}
  (\bibinfo{year}{2014}).

\bibitem{RN1676}
\bibinfo{author}{Shu, L.}, \bibinfo{author}{Wei, X.}, \bibinfo{author}{Pang,
  T.}, \bibinfo{author}{Yao, X.} \& \bibinfo{author}{Wang, C.}
\newblock \bibinfo{title}{Symmetry of flexoelectric coefficients in crystalline
  medium}.
\newblock \emph{\bibinfo{journal}{J. Appl. Phys.}}
  \textbf{\bibinfo{volume}{110}} (\bibinfo{year}{2011}).

\bibitem{RN15}
\bibinfo{author}{Li, J.}, \bibinfo{author}{Li, J.-F.}, \bibinfo{author}{Yu,
  Q.}, \bibinfo{author}{Chen, Q.~N.} \& \bibinfo{author}{Xie, S.}
\newblock \bibinfo{title}{Strain-based scanning probe microscopies for
  functional materials, biological structures, and electrochemical systems}.
\newblock \emph{\bibinfo{journal}{J. Materiomics}}
  \textbf{\bibinfo{volume}{1}}, \bibinfo{pages}{3--21} (\bibinfo{year}{2015}).

\bibitem{RN1661}
\bibinfo{author}{Nguyen, T.~D.}, \bibinfo{author}{Mao, S.},
  \bibinfo{author}{Yeh, Y.~W.}, \bibinfo{author}{Purohit, P.~K.} \&
  \bibinfo{author}{McAlpine, M.~C.}
\newblock \bibinfo{title}{Nanoscale flexoelectricity}.
\newblock \emph{\bibinfo{journal}{Adv. Mater.}} \textbf{\bibinfo{volume}{25}},
  \bibinfo{pages}{946--974} (\bibinfo{year}{2013}).
\newblock \urlprefix\url{https://www.ncbi.nlm.nih.gov/pubmed/23293034}.

\bibitem{RN1686}
\bibinfo{author}{Cao, Y.} \emph{et~al.}
\newblock \bibinfo{title}{Superlattice-induced insulating states and
  valley-protected orbits in twisted bilayer graphene}.
\newblock \emph{\bibinfo{journal}{Phys. Rev. Lett.}}
  \textbf{\bibinfo{volume}{117}}, \bibinfo{pages}{116804}
  (\bibinfo{year}{2016}).
\newblock \urlprefix\url{https://www.ncbi.nlm.nih.gov/pubmed/27661712}.

\bibitem{RN1657}
\bibinfo{author}{Kim, K.} \emph{et~al.}
\newblock \bibinfo{title}{van der waals heterostructures with high accuracy
  rotational alignment}.
\newblock \emph{\bibinfo{journal}{Nano Lett.}} \textbf{\bibinfo{volume}{16}},
  \bibinfo{pages}{1989--1995} (\bibinfo{year}{2016}).
\newblock \urlprefix\url{https://www.ncbi.nlm.nih.gov/pubmed/26859527}.

\bibitem{RN110}
\bibinfo{author}{Gruverman, A.} \& \bibinfo{author}{Kalinin, S.~V.}
\newblock \bibinfo{title}{Piezoresponse force microscopy and recent advances in
  nanoscale studies of ferroelectrics}.
\newblock \emph{\bibinfo{journal}{J. Mater. Sci.}}
  \textbf{\bibinfo{volume}{41}}, \bibinfo{pages}{107--116}
  (\bibinfo{year}{2006}).

\bibitem{RN2064}
\bibinfo{author}{Nasr~Esfahani, E.}, \bibinfo{author}{Li, T.},
  \bibinfo{author}{Huang, B.}, \bibinfo{author}{Xu, X.} \& \bibinfo{author}{Li,
  J.}
\newblock \bibinfo{title}{Piezoelectricity of atomically thin wse2 via
  laterally excited scanning probe microscopy}.
\newblock \emph{\bibinfo{journal}{Nano Energy}} \textbf{\bibinfo{volume}{52}},
  \bibinfo{pages}{117--122} (\bibinfo{year}{2018}).
\newblock
  \urlprefix\url{https://www.sciencedirect.com/science/article/pii/S2211285518305391}.

\bibitem{RN115}
\bibinfo{author}{Seol, D.}, \bibinfo{author}{Kim, B.} \& \bibinfo{author}{Kim,
  Y.}
\newblock \bibinfo{title}{Non-piezoelectric effects in piezoresponse force
  microscopy}.
\newblock \emph{\bibinfo{journal}{Curr. Appl. Phys.}}
  \textbf{\bibinfo{volume}{17}}, \bibinfo{pages}{661--674}
  (\bibinfo{year}{2017}).

\bibitem{RN1690}
\bibinfo{author}{Kresse, G.} \& \bibinfo{author}{Furthmiiller, J.}
\newblock \bibinfo{title}{Efficiency of ab-initio total energy calculations for
  metals and semiconductors using a plane-wave basis set}.
\newblock \emph{\bibinfo{journal}{Comput. Mater. Sci.}}
  \textbf{\bibinfo{volume}{6}}, \bibinfo{pages}{15--50} (\bibinfo{year}{1996}).

\bibitem{RN1274}
\bibinfo{author}{Kresse, G.} \& \bibinfo{author}{Furthmüller, J.}
\newblock \bibinfo{title}{Efficient iterative schemes for ab initio
  total-energy calculations using a plane-wave basis set}.
\newblock \emph{\bibinfo{journal}{Phys. Rev. B}} \textbf{\bibinfo{volume}{54}},
  \bibinfo{pages}{11169--11186} (\bibinfo{year}{1996}).

\bibitem{RN1271}
\bibinfo{author}{Blöchl, P.~E.}
\newblock \bibinfo{title}{Projector augmented-wave method}.
\newblock \emph{\bibinfo{journal}{Phys. Rev. B}} \textbf{\bibinfo{volume}{50}},
  \bibinfo{pages}{17953--17979} (\bibinfo{year}{1994}).

\bibitem{RN1275}
\bibinfo{author}{Kresse, G.} \& \bibinfo{author}{Joubert, D.}
\newblock \bibinfo{title}{From ultrasoft pseudopotentials to the projector
  augmented-wave method}.
\newblock \emph{\bibinfo{journal}{Phys. Rev. B}} \textbf{\bibinfo{volume}{59}},
  \bibinfo{pages}{1758--1775} (\bibinfo{year}{1999}).

\bibitem{RN1276}
\bibinfo{author}{Perdew, J.~P.}, \bibinfo{author}{Burke, K.} \&
  \bibinfo{author}{Ernzerhof, M.}
\newblock \bibinfo{title}{Generalized gradient approximation made simple}.
\newblock \emph{\bibinfo{journal}{Phys. Rev. Lett.}}
  \textbf{\bibinfo{volume}{77}}, \bibinfo{pages}{3865--3868}
  (\bibinfo{year}{1996}).

\bibitem{RN1691}
\bibinfo{author}{Grimme, S.}, \bibinfo{author}{Antony, J.},
  \bibinfo{author}{Ehrlich, S.} \& \bibinfo{author}{Krieg, H.}
\newblock \bibinfo{title}{A consistent and accurate ab initio parametrization
  of density functional dispersion correction (dft-d) for the 94 elements
  h-pu}.
\newblock \emph{\bibinfo{journal}{J. Chem. Phys.}}
  \textbf{\bibinfo{volume}{132}}, \bibinfo{pages}{154104}
  (\bibinfo{year}{2010}).
\newblock \urlprefix\url{https://www.ncbi.nlm.nih.gov/pubmed/20423165}.

\end{thebibliography}

\clearpage

\renewcommand{\thefigure}{S\arabic{figure}}
\renewcommand{\thesection}{S\arabic{section}}
\renewcommand{\thesubsection}{S\arabic{subsection}}
\renewcommand{\theequation}{S\arabic{equation}}
\renewcommand{\thetable}{S\arabic{table}}
\setcounter{figure}{0} 
\setcounter{equation}{0}
\appendix 

\onecolumngrid

\section*{Supplementary Information}

\begin{figure*}[htbp]	  
	\centering
	\includegraphics[width=6in] {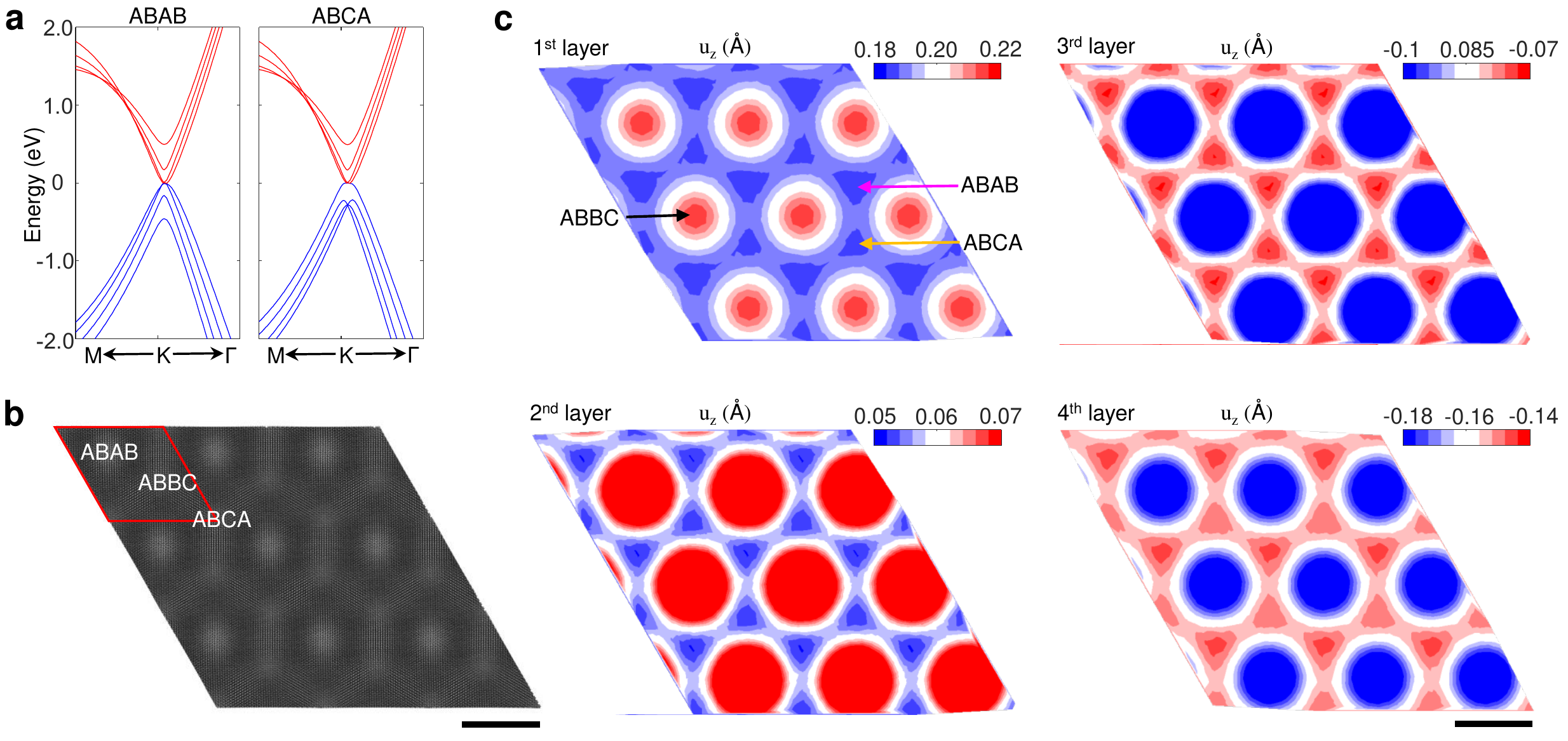}
	\caption
	{ \textbf{DFT calculation of tDBG.}
		\textbf{(a)} Band structures of ABAB and ABCA stacking graphene.
		\textbf{(b)} Modeling of tDBG moiré superlattice in DFT calculation. Scale bar, 2 nm.
		\textbf{(c)} Out-of-plane displacement ($u_z$) fields of four graphene layers in the reconstructed tDBG moiré superlattice. The ABAB, ABCA and ABBC regions are pointed out with pink, orange and black arrow, respectively. Scale bar, 2 nm.
	}
	\label{fig:S1} 
\end{figure*}

\begin{figure*}[htbp]	  
	\centering
	\includegraphics[width=6in] {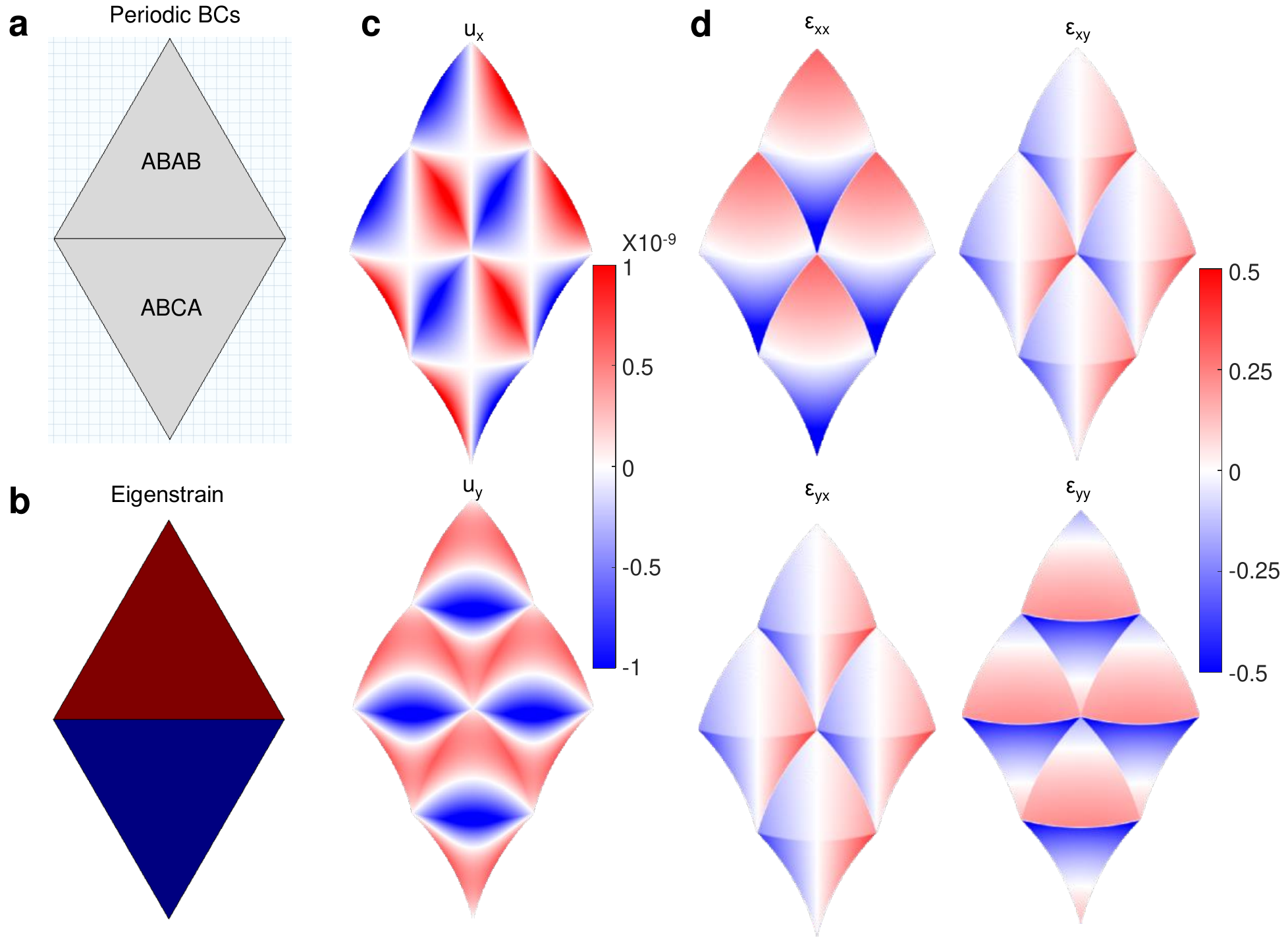}
	\caption
	{ \textbf{Continuum mechanics simulation of tDBG.}
		\textbf{(a)} Modeling of tDBG unit cell with periodic boundary condition (BC).
		\textbf{(b)} Eigenstrain in ABAB and ABCA domains.
		\textbf{(c)} Distribution of displacements ($u_x$ and $u_y$) with deformation.
		\textbf{(d)} Distribution of four strain components ($\varepsilon_{xx}$, $\varepsilon_{xy}$, $\varepsilon_{yx}$ and $\varepsilon_{yy}$) with deformation.
	}
	\label{fig:S2} 
\end{figure*}

\begin{figure*}[htbp]	  
	\centering
	\includegraphics[width=6in] {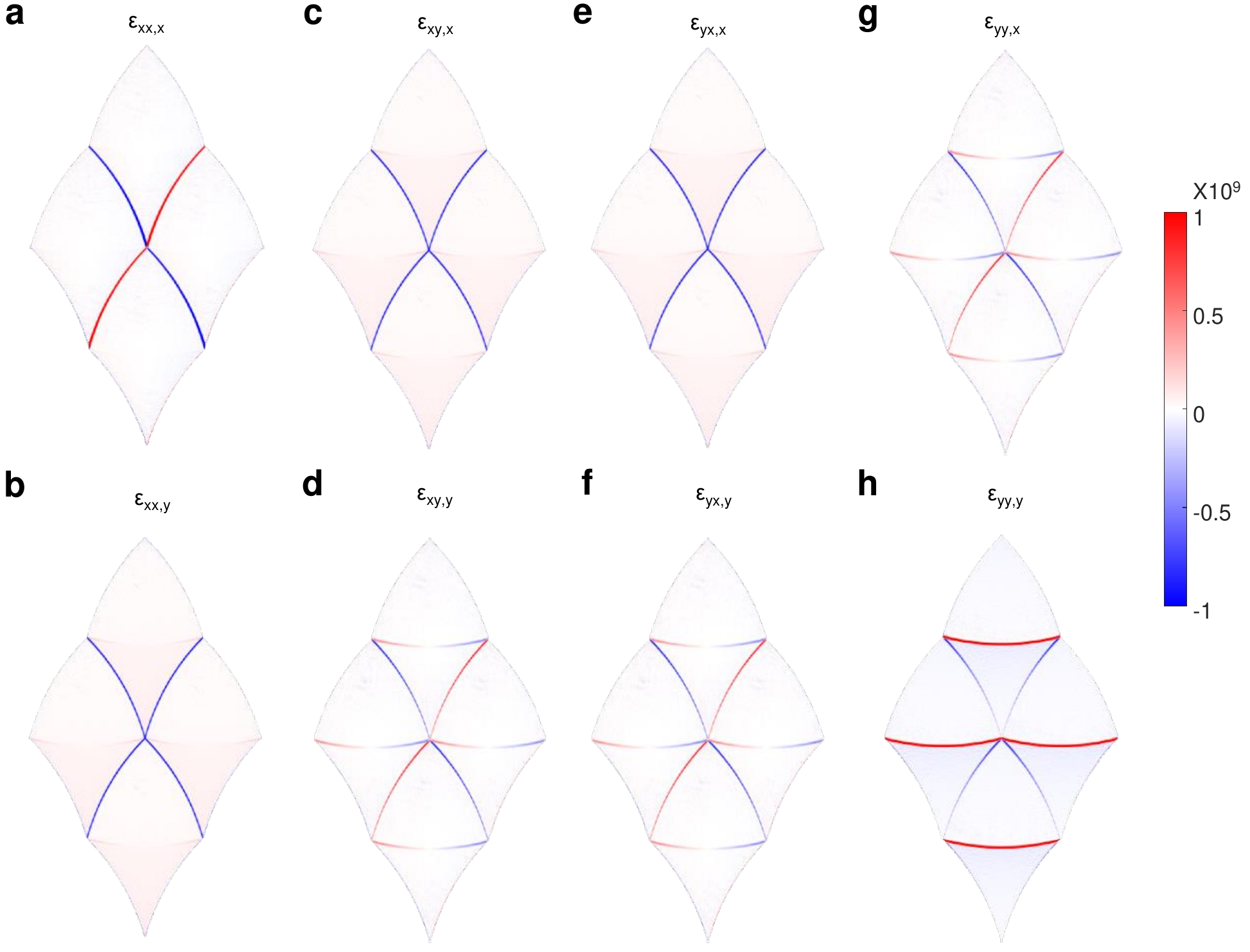}
	\caption
	{ \textbf{Continuum mechanics simulation of tDBG.}
		\textbf{(a-h)}  Distribution of eight strain gradient components with deformation.
	}
	\label{fig:S3} 
\end{figure*}

\begin{figure*}[htbp]	  
	\centering
	\includegraphics[width=6in] {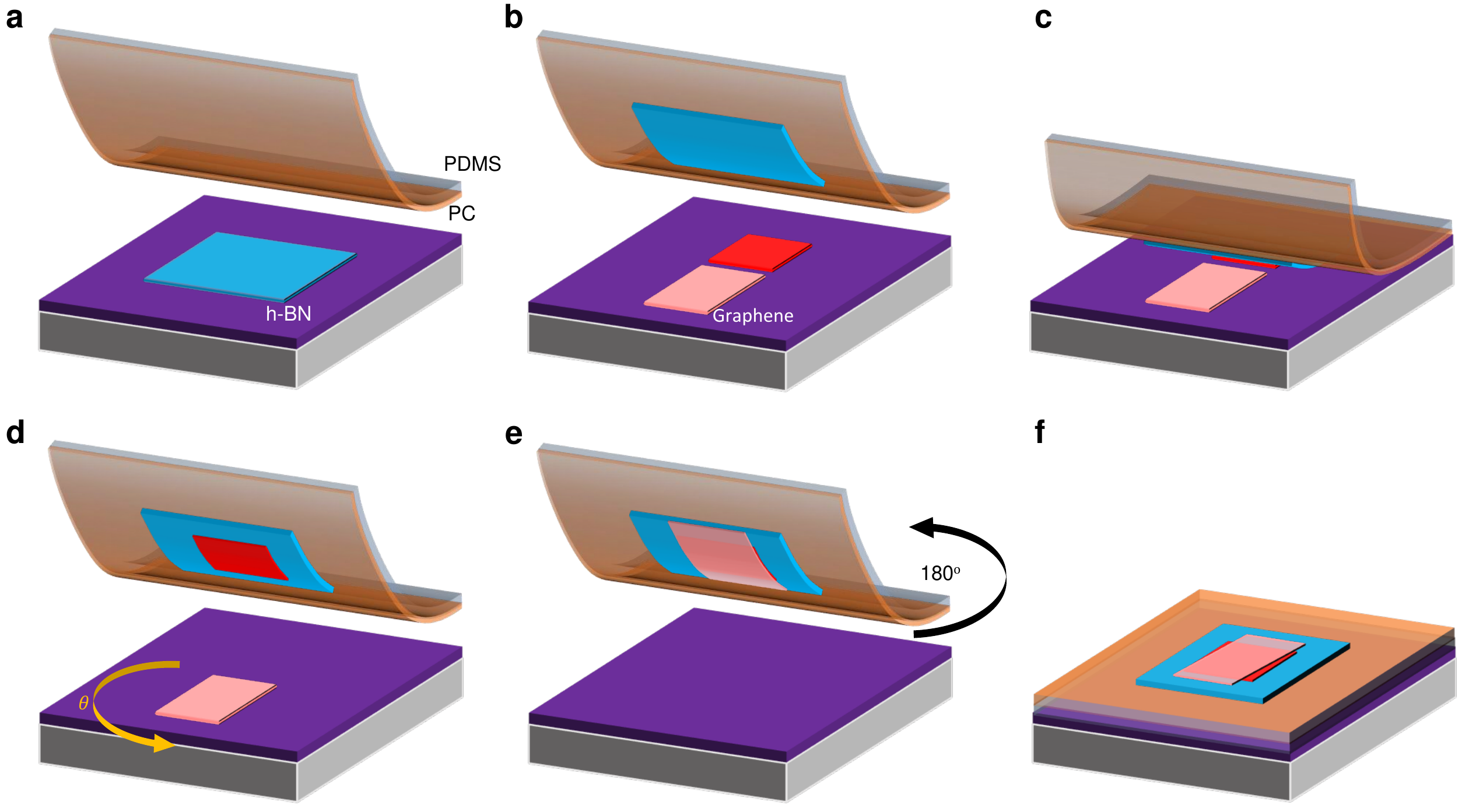}
	\caption
	{ \textbf{Processes of sample fabrication.}
		\textbf{(a)} A PC film on top of a PDMS stamp is used to pick up an h-BN-layer.
		\textbf{(b,c)} Using the relatively big adhesive force between h-BN and graphene, half of a cut bilayer graphene is picked up in a low temperature ($\sim$40~$^{\circ}$C).
		\textbf{(d,e)}  By rotating the sample to a desired angle, the second half of bilayer graphene is picked up, and the transfer process is finished.
		\textbf{(f)} After transferring, the PC film is flipped over and melted onto a Si/SiO$_2$ substrate.
	}
	\label{fig:S4} 
\end{figure*}

\begin{figure*}[htbp]	  
	\centering
	\includegraphics[width=6in] {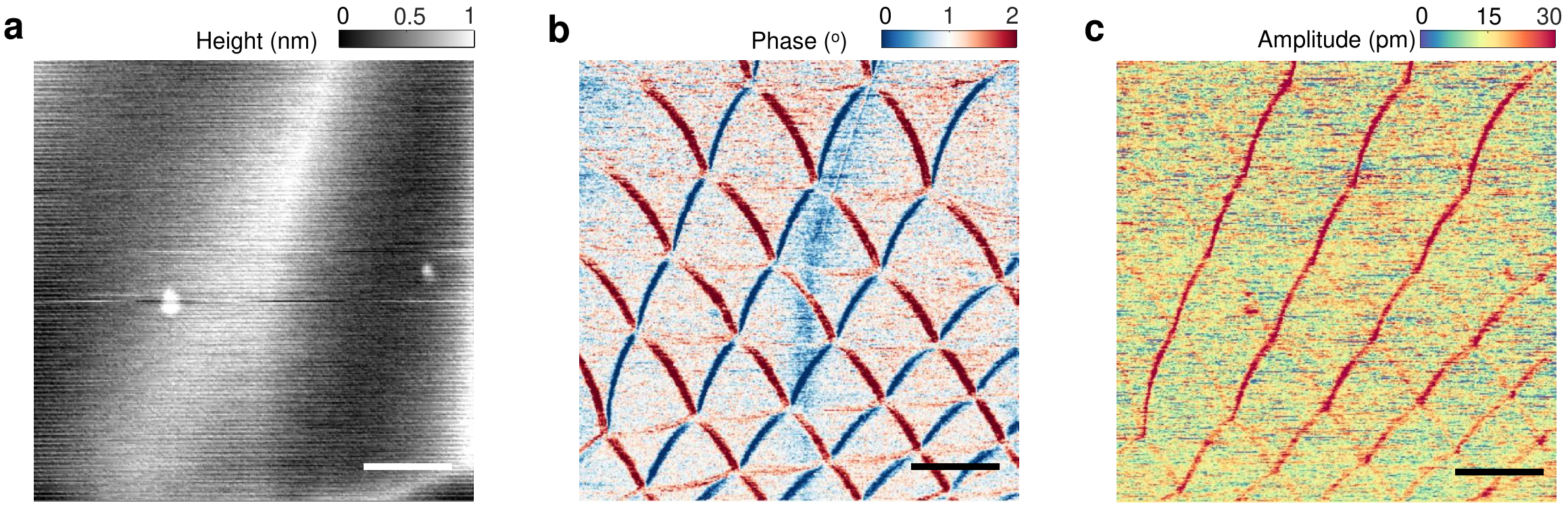}
	\caption
	{ \textbf{Flattened LPFM results of tDBG.}
		\textbf{(a-c)} Flattened height (a), phase (b) and amplitude (c) of LPFM results. The direction of cantilever is parallel to the horizontal. Scale bars, 400 nm.
	}
	\label{fig:S5} 
\end{figure*}

\begin{figure*}[htbp]	  
	\centering
	\includegraphics[width=4in] {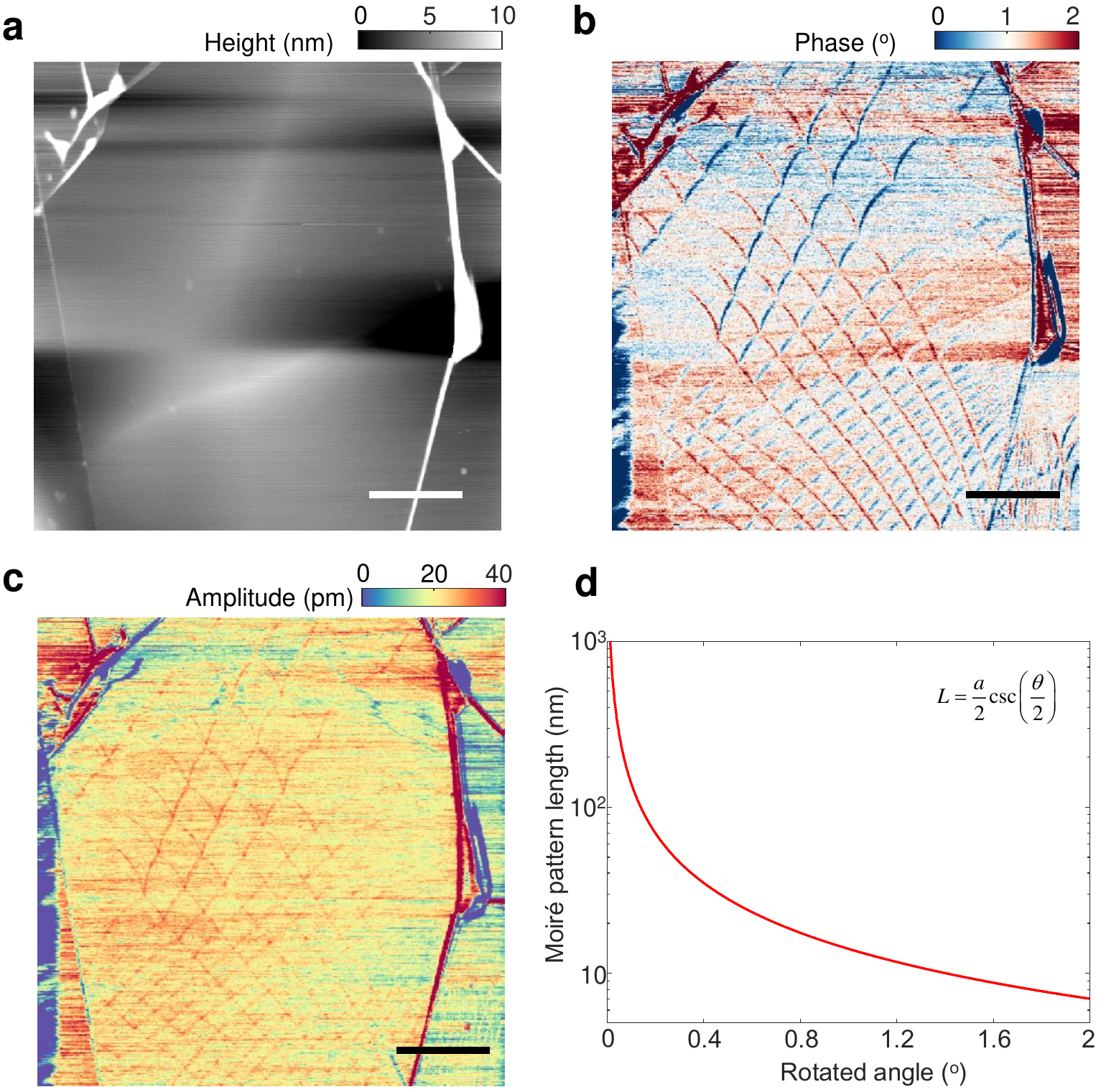}
	\caption
	{ \textbf{Flattened large scan LPFM results in tDBG.}
		\textbf{(a-c)} Height (a), phase (b) and amplitude (c) in a large LPFM scan. Scale bars, 1 $\mu$m.
		\textbf{(d)} Relation between rotated angle ($\theta$) and moiré pattern length ($L$). The direction of cantilever is parallel to the horizontal. 
	}
	\label{fig:S6} 
\end{figure*}

\begin{figure*}[htbp]	  
	\centering
	\includegraphics[width=6in] {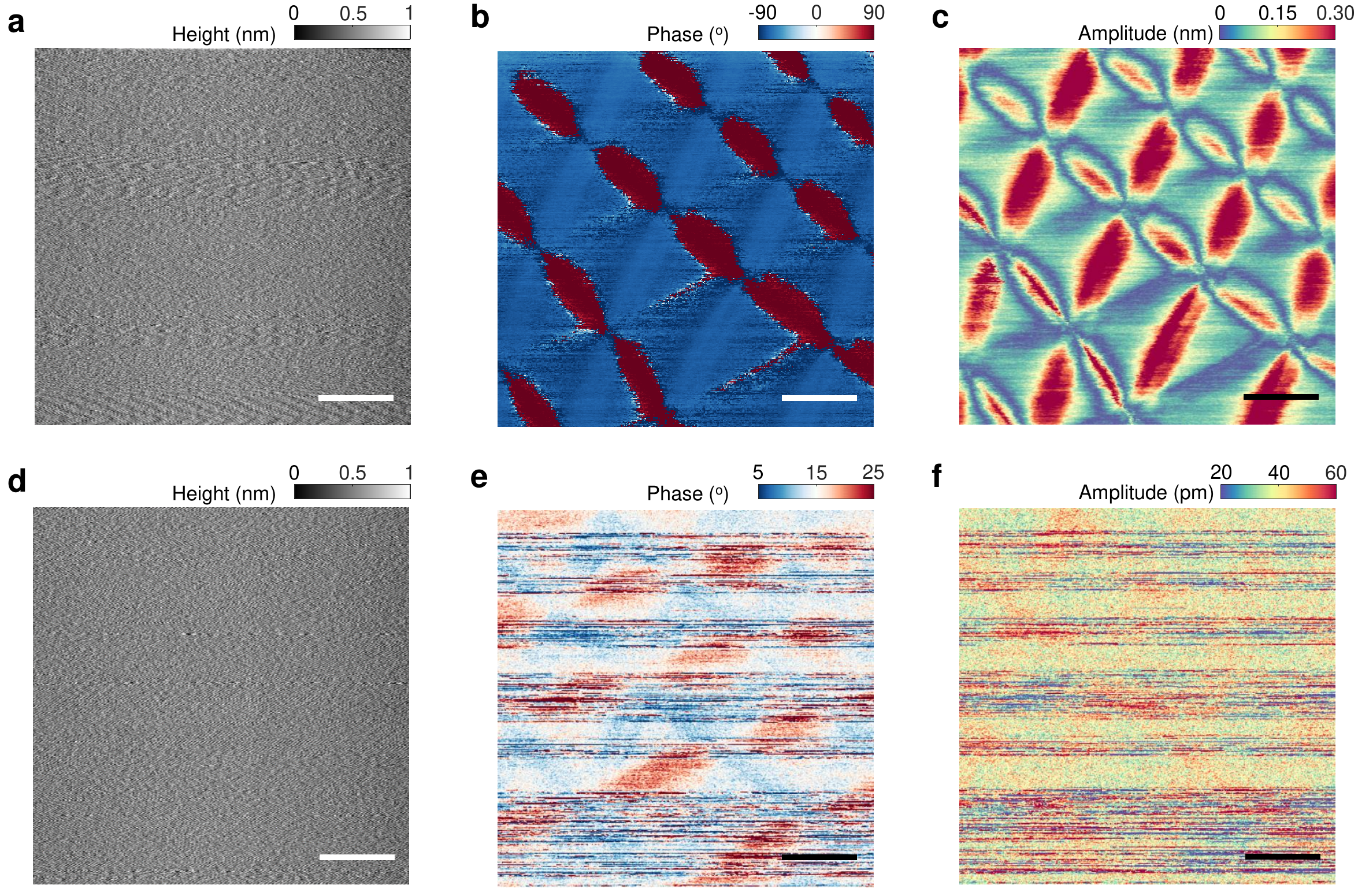}
	\caption
	{ \textbf{In-situ LPFM and VPFM of tDBG.}
		\textbf{(a-c)} LPFM results of tDBG. The direction of cantilever is parallel to the horizontal. Scale bars, 100 nm.
		\textbf{(d-e)} VPFM results of tDBG. The direction of cantilever is parallel to the horizontal. Scale bars, 100 nm.
	}
	\label{fig:S7} 
\end{figure*}

\begin{figure*}[htbp]	  
	\centering
	\includegraphics[width=\textwidth] {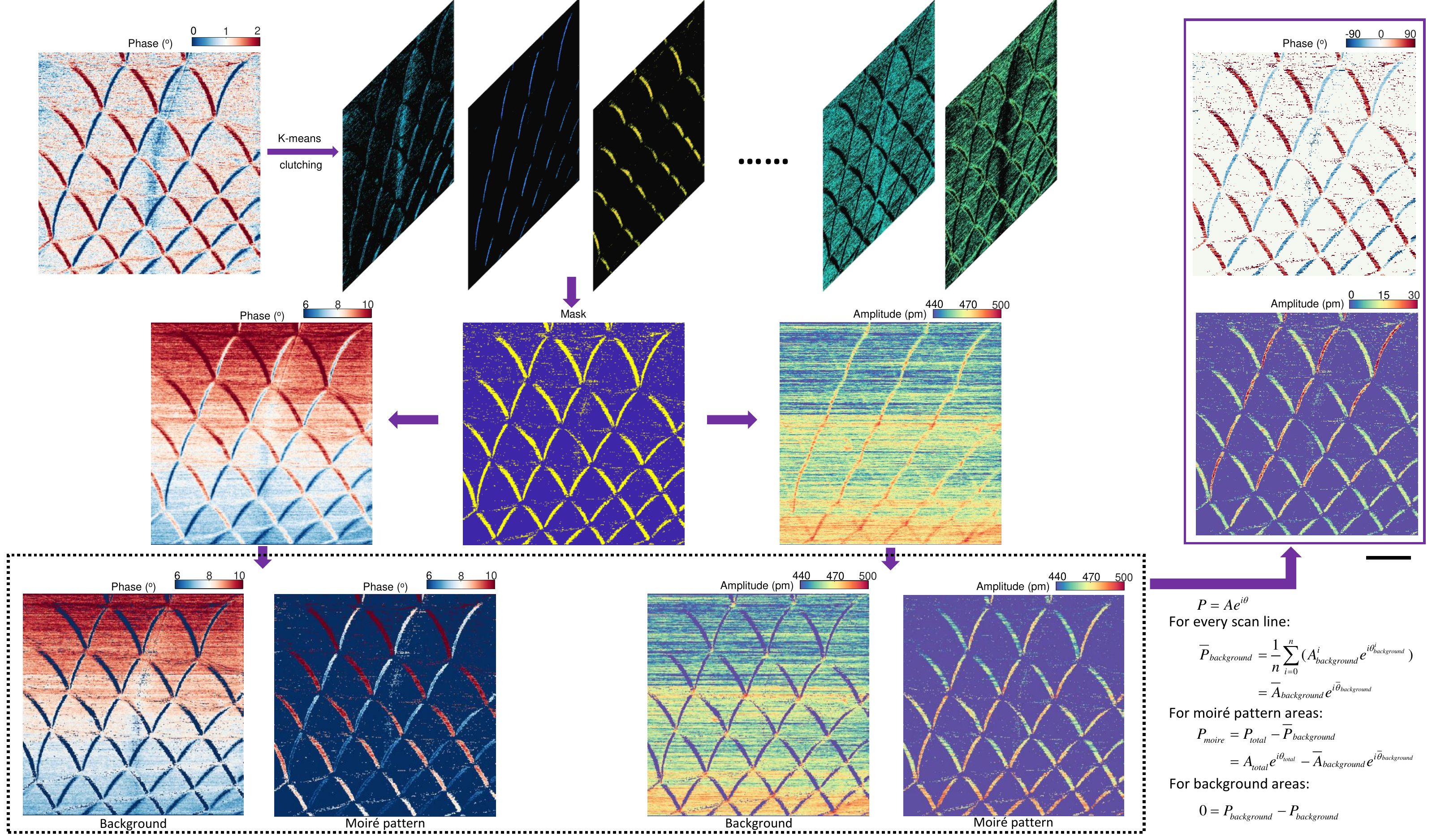}
	\caption
	{ \textbf{Processes of decoupling LPFM results.}
		Firstly, the flatten phase is clutched using K-means clutching, the phase image can be decomposed into serval images. By choosing the image only containing DWs signals, a mask is determined. By applying this mask to the raw data of LPFM, the background and moiré pattern are distinguished. Because of the variable background signal, the background signal is determined for every scan line. By subtracting the background vector for every moiré pattern point and setting every background point zero, the pure moiré pattern LPFM mapping is determined. Scale bar, 400 nm.
	}
	\label{fig:S8} 
\end{figure*}

\begin{figure*}[htbp]	  
	\centering
	\includegraphics[width=6in] {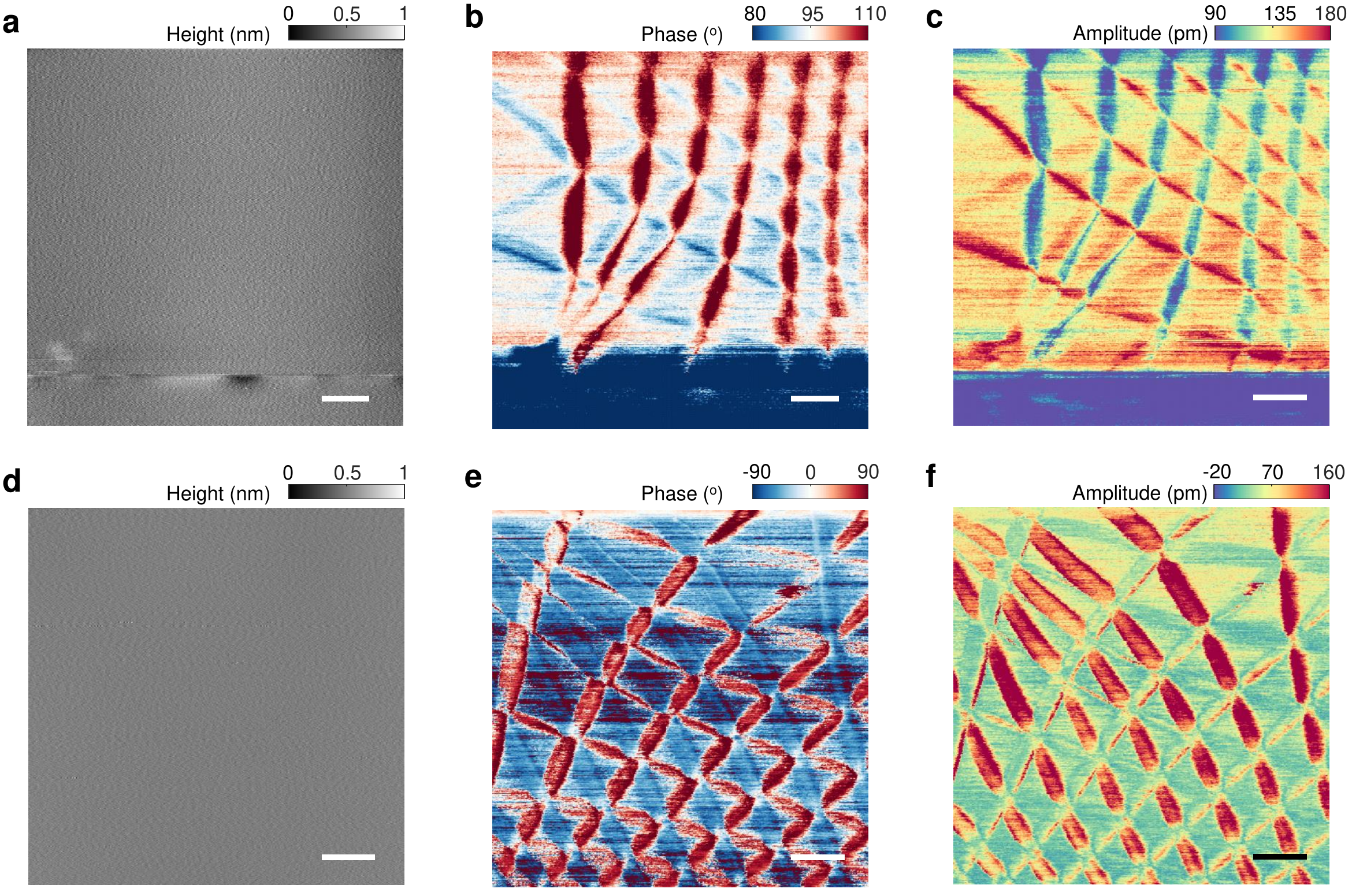}
	\caption
	{ \textbf{Raw LPFM results of tDBG at 0 and 90 degrees.}
		\textbf{(a-c)} LPFM results of tDBG at 0 degree. The direction of cantilever is perpendicular to the horizontal. Scale bars, 100 nm.
		\textbf{(d-f)} LPFM results of tDBG at $\sim$90 degree. The direction of cantilever is parallel to the horizontal. Scale bars, 100 nm.
	}
	\label{fig:S9} 
\end{figure*}

\end{document}